\documentclass[namedreferences]{solarphysics}
\usepackage[hyperref,optionalrh,solaromanenum]{spr-sola-addons} 
\usepackage{natbib}                     
\usepackage{graphicx}                    
\usepackage{color}                       
\usepackage{breakurl}                         

\chardef\us=`\_

\begin{document}

\begin{article}

\begin{opening}

\title{The Grad-Shafranov Reconstruction of Toroidal Magnetic Flux Ropes: Method Development and Benchmark Studies}

%
 \author[addressref={aff1},email={qiang.hu.th@dartmouth.edu}]{\inits{Q.~}\fnm{Qiang}~\lnm{Hu}\orcid{0000-0002-7570-2301}}

%
\runningauthor{HU} \runningtitle{GS Reconstruction in Toroidal
Geometry}

\address[id={aff1}]{Department of Space Science and CSPAR, The University of Alabama in Huntsville, Huntsville, AL 35805}

\begin{abstract}
We develop an approach of Grad-Shafranov (GS) reconstruction for
toroidal structures in space plasmas, based on in-situ spacecraft
measurements. The underlying theory is the GS equation that
describes two-dimensional magnetohydrostatic equilibrium as widely
applied in fusion plasmas. The geometry is such that the arbitrary
cross section of the torus has rotational symmetry about the
rotation axis $Z$, with a major radius $r_0$. The magnetic field
configuration is thus determined by a scalar flux function $\Psi$
and a functional $F$ that is a single-variable function of $\Psi$.
The algorithm is implemented through a two-step approach: i) a
trial-and-error process by minimizing the residue of the
functional $F(\Psi)$ to determine an optimal $Z$ axis orientation,
and ii) for the chosen $Z$, a $\chi^2$ minimization process
resulting in the range of $r_0$. Benchmark studies of known
analytic solutions to the toroidal GS equation with noise
additions are presented to illustrate the two-step procedures and
to demonstrate the performance of the numerical GS solver,
separately. For the cases presented, the errors in $Z$ and $r_0$
are 9$^\circ$ and 22\%, respectively, and the relative percent
error in the numerical GS solutions is less than 10\%. We also
make public the computer codes for these implementations and
benchmark studies.
\end{abstract}

%


\end{opening}

%
 \section{Introduction}\label{s:intro}
Magnetic flux rope modeling based on in-situ spacecraft
measurements plays a critical role in characterizing this type of
magnetic and plasma structures. Simply put, it provides the most
direct, definitive and quantitative evidence for the existence of
such structures and their characteristic configuration of that of
a magnetic flux rope with winding magnetic field lines embedded in
space plasmas of largely magnetohydrostatic equilibrium. Such
analysis dated back to early times of the space age, especially
with the discovery of Magnetic Clouds (MCs) from in-situ solar
wind data \citep[see, e.g.,][and references therein]{1995ISAAB}.
Among these modeling methods, employing in-situ magnetic field and
plasma time-series data across such structures, the so-called
Grad-Shafranov (GS) reconstruction method stands out as one (and
the only one) truly two-dimensional (2D) method that derives the
cross section of a flux rope in complete 2D configuration, or more
precisely, $2\frac{1}{2}$D, with two transverse magnetic field
components lying on the cross-sectional plane and the
non-vanishing axial component perpendicular to the plane.

The conventional GS method applies to a flux rope configuration of
translation symmetry, i.e., that of a straight cylinder with a
fixed axis, but of an arbitrary 2D cross section perpendicular to
it. Therefore the field lines are winding along such a central
axis lying on distinct and nested flux surfaces  defined by an
usual flux function in 2D geometry. The GS method for a
straight-cylinder geometry was first proposed by
\citet{1996GeoRLS}, later further developed to its present form by
\citet{1999JGRH} and applied to magnetopause current sheet
crossings \citep[see also,][]{2000GeoRLH,2003JGRAH}. It was first
applied to the flux rope structures in the solar wind by
\citet{2001GeoRLHu}, at first to the small-scale ones of durations
$\sim$30 minutes, then to the large-scale MCs with detailed
descriptions of the procedures tailored toward this type of GS
reconstruction in \citet{2002JGRAHu}. Since then, the GS
reconstruction method has been applied to the solar wind in-situ
measurements of MCs by a number of research groups
\citep[e.g.,][]{2016ApJ...829...97H,2016ApJ...828...12V,2016JGRA..121.7423W,2016GeoRL..43.4816H,2013JGRA..118.3954S,2012ApJ...758...10M,2009SoPh..254..325K,2009SoPh..256..427M,2009AnGeo..27.2215M,2008AnGeo..26.3139M,2008ApJ...677L.133L,2007JGRA..112.9101D}.
For a detailed review of the works related to GS reconstruction of
magnetic flux rope structures, see \citet{Hu2017GSreview}.

 Challenges facing the in-situ flux rope modeling including
GS reconstruction stem from the variabilities in the
configuration, properties and origins of magnetic flux ropes,
concerning the MCs. For example, \citet{2011JGRAK} examined a number
of MC events at 1 AU, interpreted as magnetic flux ropes using
relatively simple  models of axi-symmetric cylindrical
configuration. By comparing directly the modeled field-line
lengths with the ones measured by traversing energetic electrons
from the Sun to 1 AU  \citep{1997GeoRLL}, they concluded that
the MC flux rope configuration, interpreted by the commonly
known linear force-free model \citep{lund}, is not consistent with such measurements.  On the other hand, we
showed in \citet{2015JGRAH} that for the same set of measurements,
the field-line length estimates from the GS reconstruction results
agree better with such measured path lengths from electron burst
onset analysis. In addition to these unique measurements for the purpose of
validating flux rope models, we also attempted indirect means by
relating the in-situ GS flux rope model outputs with the corresponding solar source
region properties. In an early work \citep{Qiu2007}, we established certain
correlation between the magnetic flux contents and the corresponding flare
reconnection flux on the Sun. Following that work,
\citet{2014ApJH} further extended the analysis to derive magnetic
field-line twist distributions inside MCs based on GS
reconstruction results, and hinted at the formation mechanism of
flux ropes, at least partially, due to morphology in flares or
magnetic reconnection sequence, thus leading to the variability in
the twist distributions as observed from in-situ data.
Capitalizing on these findings based on both in-situ flux rope
modeling and observational analysis on the Sun, theoretical
investigations \citep{2016SoPh..291.2017P,2016SoPhP2} were also
attempted very recently to probe the formation of flux ropes due
to magnetic reconnection, as manifested by solar flares.
 Therefore, it is imperative to {\textbf{further develop the}}
existing approaches of flux rope modeling to account for such
variabilities in order to shed light on the important question
regarding the origination and formation of magnetic flux ropes from the Sun.

In the present study, we intend to address the variability
concerning the configuration of a magnetic flux rope, by extending
the applicability of the GS reconstruction method to the geometry
of a torus. 
{\textbf{We acknowledge that such an extension is not meant to be
a replacement of the cylindrical flux-rope model, but an addition
or an alternative to the toolset of flux rope modeling. The
advantage of such a configuration over a straight-cylinder has to
be assessed on a case-by-case basis. Sometimes it offers a useful
and complementary alternative to the straight-cylinder model,
especially when the latter model fails (see, e.g.,
Section~\ref{subsec:solver}).}}

A word of caution is that we only use a section of the torus to
approximate the local structure of the flux rope in the vicinity
of the spacecraft path across the toroidal section. Otherwise it
would have implied that the flux ropes as detected in-situ possess
a closed configuration with complete detachment from the Sun which
has generally been refuted \citep[e.g.,][]{1995ISAAB}. However, a
number of numerical simulations have utilized a closed magnetic
configuration similar to that of a typical tokamak (or spheromak)
to initiate CMEs close to the Sun
\citep[e.g.,][]{2016SpWea..14...56S}. In fusion sciences, the
confined plasma experiments always have a closed geometry, e.g., a
tokamak of axi-symmetric toroidal configuration
\citep{freidberg87}. In this study, we try to tap into the wealth
of knowledge in fusion plasma science describing 2D configurations
in ideal magnetohydrodynamic (MHD) equilibria under such a
geometry.


Somewhat as done before \citep[see, e.g.,][]{2008JGRAS,2009JGRAS},
we adopt the practice  of presenting basic theoretical
consideration, analysis procedures and benchmark studies first in
this presentation, but leave some more comprehensive benchmark
studies and application to real events to a follow-up publication.
This serves the purpose of not overwhelming the reader and
ourselves, but guaranteeing a relatively short and focused report
of the new development of this technique to benefit the user
community.

The article is organized as follows. The GS equation in the
toroidal geometry and the basic setup of the reconstruction frame
are described in Section~\ref{sec:GSeq}. Then a recipe in terms of
a two-step reconstruction procedure is described in detail in
Section~\ref{sec:proc}. Benchmark studies of the basic procedures
and the performance of the numerical GS solver are given in
Section~\ref{sec:bench}. We conclude in the last section, followed
by several appendices laying out additional details and a special
situation to be considered. We emphasize that the focus of this
article is to allow interested readers to perform their own case
studies and to devise their own computer codes if they choose to,
facilitated by the detailed descriptions and the auxiliary
material including the complete set of computer codes implemented
in Matlab.

\section{Grad-Shafranov Equation in Toroidal
Geometry}\label{sec:GSeq}

Equivalent to the GS equation in a Cartesian geometry on which the
traditional GS reconstruction method is based, there is a GS
equation in the so-called toroidal geometry of rotational
symmetry, given in a usual cylindrical coordinate $(R,\phi,Z)$:
\begin{equation}
R\frac{\partial}{\partial
R}\left(\frac{1}{R}\frac{\partial\Psi}{\partial
R}\right)+\frac{\partial^2\Psi}{\partial
Z^2}=-\mu_0R^2\frac{dp}{d\Psi}-F\frac{dF}{d\Psi}.\label{eq:GSt}
\end{equation}
\begin{figure}
 \centerline{\includegraphics[width=0.6\textwidth,clip=]{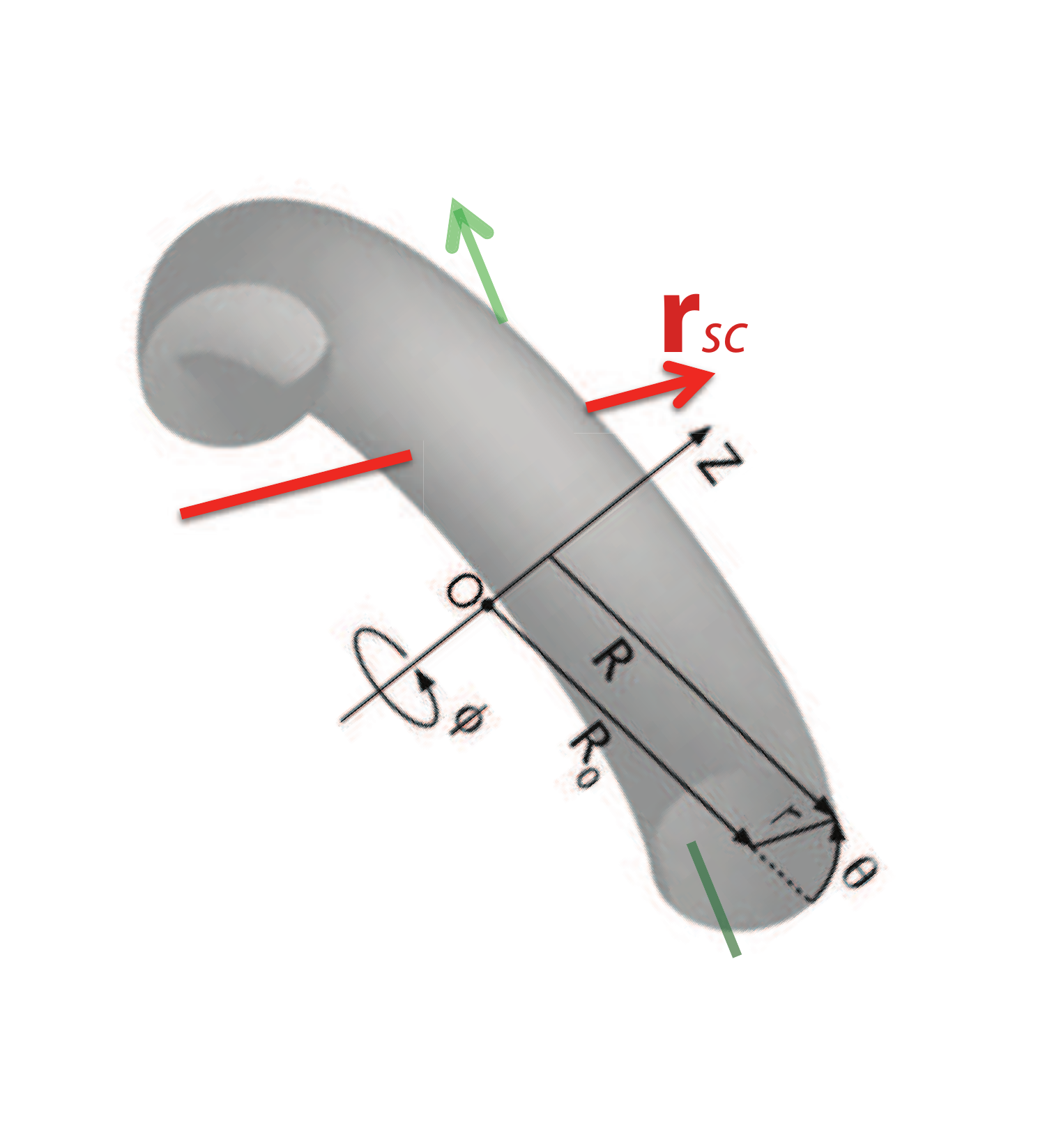}}
 \caption{Illustration of the toroidal geometry (a section of the torus) with respect to the spacecraft path
 $\mathbf{r}_{sc}$ (adapted from FusionWiki, {\url{http://fusionwiki.ciemat.es/wiki/Toroidal_coordinates}}).
 The cross section of the torus is either described  in the cylindrical coordinate $(R,\phi,Z)$ or
 spherical coordinate $(r,\theta,-\phi)$. {\textbf{Here the parameter $R_0$ is   chosen such that
 $R_0=r_0$, the major radius. See Figure~\ref{fig:RSZ} for the actual coordinate systems and a different
 choice of $R_0$ utilized throughout this study.}} The relation $R=R_0+r\cos\theta$ always satisfies. The other straight
  line in green illustrates a submerged
 path which cannot be included in the current toroidal GS reconstruction approach,
 but is discussed in the Appendix~\ref{app:hodo}. }
 \label{fig:torcoord}
 \end{figure}
As illustrated in Figure~\ref{fig:torcoord}, the above GS equation
describes the space plasma structure in quasi-static equilibrium
of rotational symmetry, i.e., that of a torus. The configuration
is fully characterized by a cross section of such a torus rotating
around the rotation axis, $Z$, thus yielding invariance in the
azimuthal $\phi$ direction, i.e., $\partial/\partial\phi\approx
0$. Under this geometry, the magnetic field vector is
\begin{equation} \mathbf{B}=\frac{1}{R}\nabla\Psi\times\hat{\mathbf{e}}_\phi+
\frac{F(\Psi)}{R}\hat{\mathbf{e}}_\phi,\label{eq:B} \end{equation}
where the (poloidal) flux function $\Psi$ characterizes the
transverse field components and has the unit of Wb/radian. The
plasma pressure $p$ and the composite function $F=RB_\phi$,
appearing in the right-hand side of equation~(\ref{eq:GSt}),
become functions of $\Psi$ only. Therefore  similar to the
straight-cylinder case, the 2D magnetic field components plus the
out-of-plane one ($B_\phi$) are derived from the spacecraft
measurements along its path across (along $-{\mathbf{r}}_{sc}$ in
Figure~\ref{fig:torcoord}) by solving the toroidal GS
equation~(\ref{eq:GSt}) over certain cross-sectional domain. In
practice, the numerical GS solver is implemented for the GS
equation written in the alternative $(r,\theta)$ coordinate
\citep{freidberg87}:
\begin{equation}
\frac{1}{r}\frac{\partial}{\partial
r}\left(r\frac{\partial\Psi}{\partial r}\right)
+\frac{1}{r^2}\frac{\partial^2\Psi}{\partial
\theta^2}-\frac{1}{R}\left(\cos\theta\frac{\partial\Psi}{\partial
r}-\frac{\sin\theta}{r}\frac{\partial\Psi}{\partial\theta}\right)=-\mu_0
R^2\frac{dp}{d\Psi}-F\frac{dF}{d\Psi}.\label{eq:GSrth}
\end{equation}

In this geometry, there are two main geometrical  parameters to be
determined, the orientation of the rotation axis $Z$, and the
major radius $r_0$, whereas in the straight-cylinder case, only
one parameter, namely the axis orientation of the cylinder, is to
be determined. We note that the major radius can be either defined
as the radial distance between the rotation axis and the
geometrical center of the cross section of the torus or the
distance to the location where the poloidal (transverse) magnetic
field vanishes.
We adopt the former in this study since it is the convention for
plasma confinement studies \citep{freidberg87}. Inevitably, the
parameter space is much enlarged in the present case and the
reconstruction procedures are more evolved  as to be described in
the following section.

\section{Procedures of Toroidal GS Reconstruction}\label{sec:proc}
The procedures are presented for the most general cases of
arbitrary orientation of the $Z$ axis and a relatively wide range
of major radii of the torus. The analysis is primarily performed
in the spacecraft or Sun centered $r_{sc}tn$ coordinate system (to
distinguish from the local spherical coordinate $r,\theta,-\phi$;
see Figure~\ref{fig:RSZ}), where the radial direction is always
along the Sun-spacecraft line, assuming a radially propagating
solar wind carrying the structure.

We present a two-step recipe that is based on an extensive
benchmark study of known analytic solutions. We stress that this
is the best approach we have found so far, based on our experience
and largely empirical studies. It is our intention to present what
we have devised, deemed an optimal approach, and to deliver the
reconstruction code to the user community for a timely release,
for the purpose of much enhanced and collective effort in further
validation and application of the toroidal GS reconstruction
beyond the limitations of a solo effort. This is also the reason
for our concise presentation of a ``recipe" accompanied by the
computer codes to enable others to either repeat the results or to
generate their own.

\subsection{The Most General Case}\label{subsec:general}

\begin{figure}
 \centerline{\includegraphics[width=0.45\textwidth,clip=]{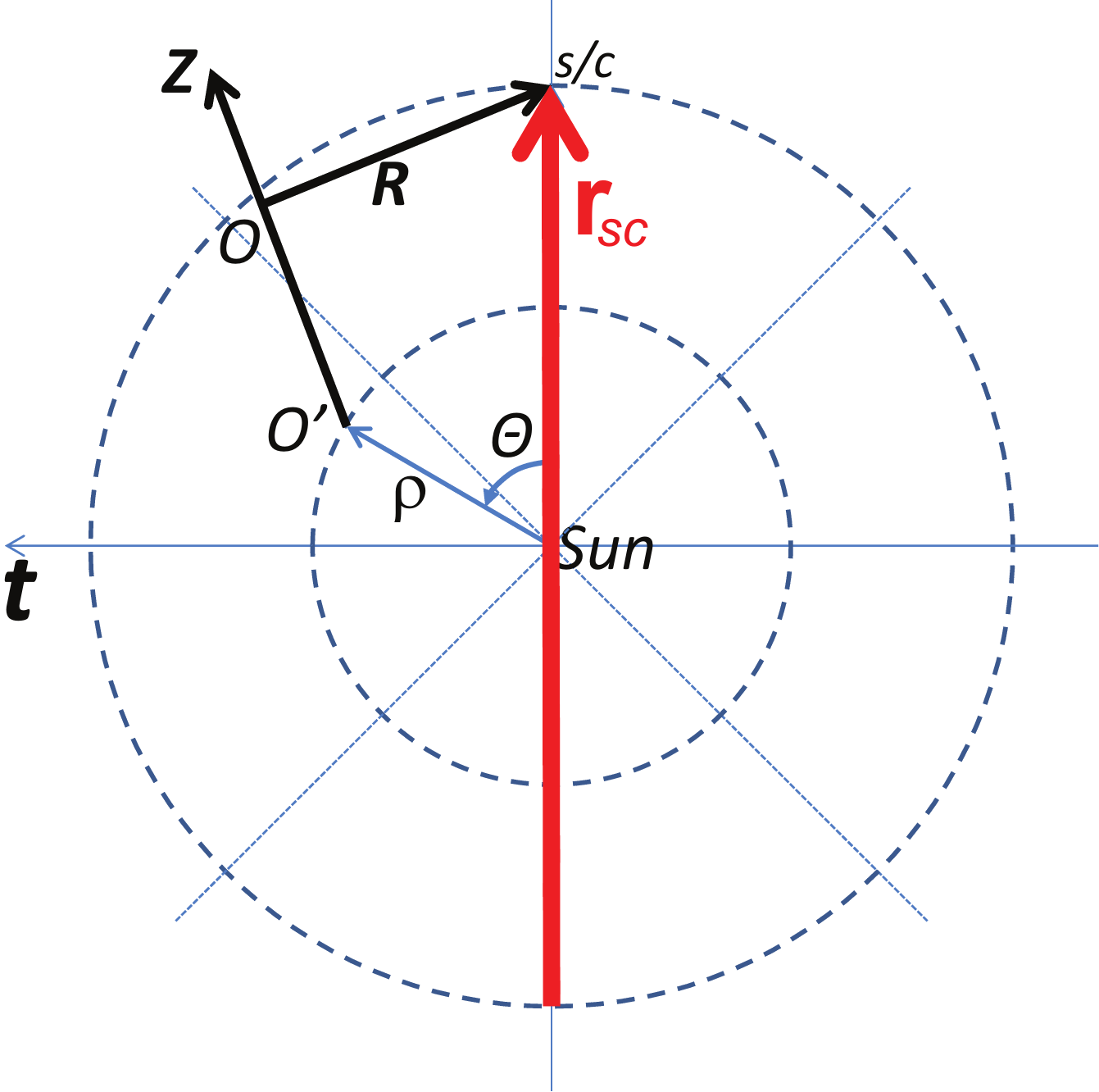}
 \includegraphics[width=0.55\textwidth,clip=]{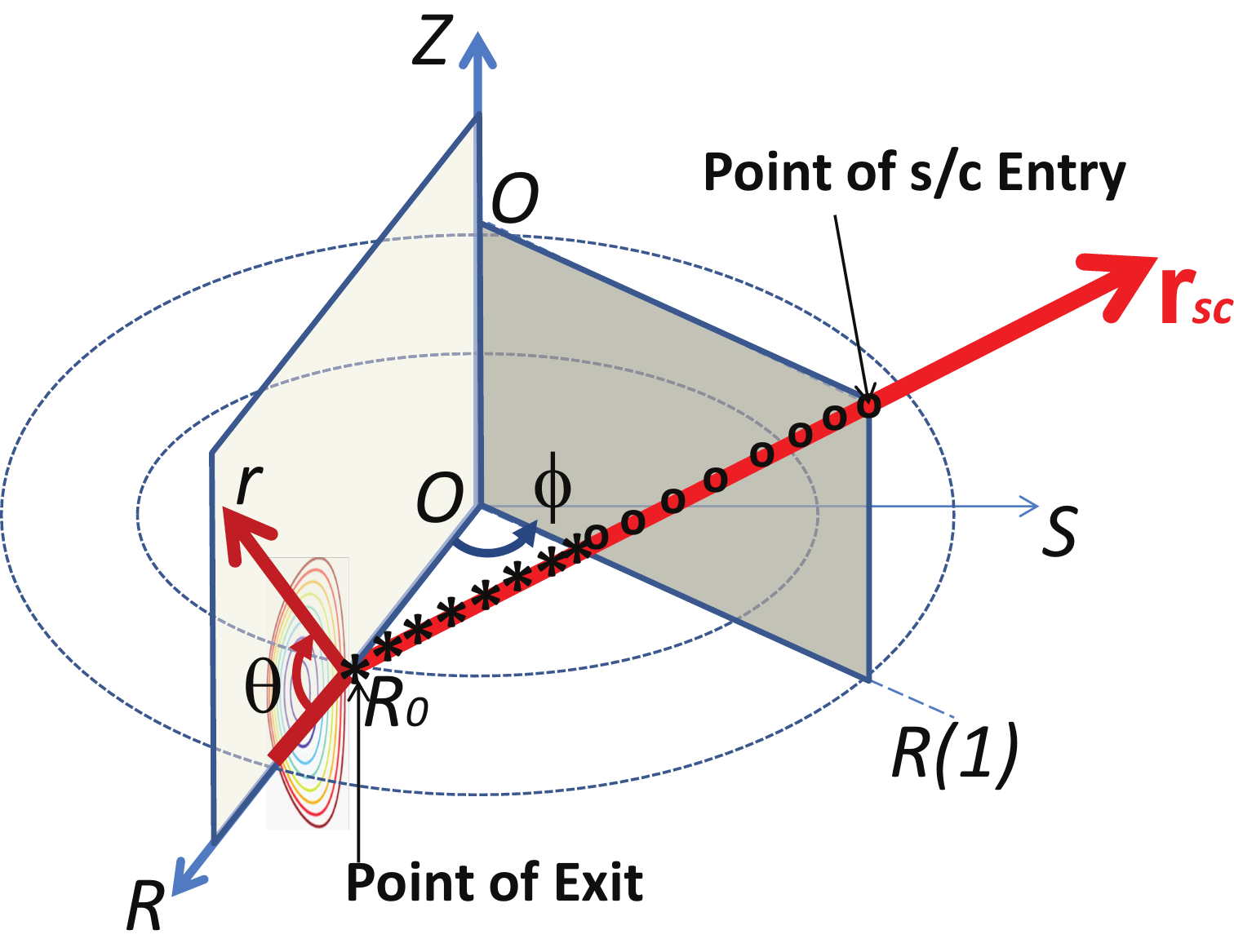}}
 \caption{(Adapted from \citet{2015ASPCH}) Left panel: the layout of search grid $O'(\rho,\Theta)$ on the ${r}_{sc}t$ plane. At each point
 $O'$, a trial-and-error process is performed for a trial $Z$ axis of arbitrary orientation in space. {\textbf{The polar coordinate $\rho$ is
 up to the outer circle of radius the radial distance of the spacecraft to the Sun. }} Right panel: the view from a different
 perspective of the local reconstruction frame $RSZ$ or $R\phi Z$ (of origins $O$), for a chosen $Z$ axis.
 {\textbf{The parameter $R_0$ is chosen as the distance from $Z$ to the point of spacecraft exit of the cross section bounded by
 the two ellipses (circles) along $R$ (at $Z$=0). The major radius, $r_0$, is then the distance from $Z$ to the middle of the two boundaries along $R$. The circles and stars represent the in-situ spacecraft data collected along the inbound
 and outbound path with respect to the point of the closest approach distance (so-called impact parameter) to the center of the flux rope structure.
 They would correspond
 to the same set of symbols in Figure~\ref{fig:pta003}. The corresponding search grid point, $O'$, along $Z$,
  is omitted in the right panel.}}
 The spacecraft path  is projected onto the light-shaded $ROZ$ plane, along $r$ of an approximately constant
 $\theta=\theta_0$. The final reconstruction of the cross section is performed on the $ROZ$ plane as illustrated. }\label{fig:RSZ}
 \end{figure}
As illustrated in Figure~\ref{fig:RSZ}, left panel, the most
general case corresponds to a torus of arbitrary major radius and
$Z$ axis orientation, whose central rotation axis intersects the
$r_{sc}t$ plane at point $O'$. Then relatively speaking, the
spacecraft is moving along $-\mathbf{r}_{sc}$ across the torus,
viewed in the frame of reference moving with the structure,
usually the deHoffmann-Teller (HT) frame (taking the radial component only) that is well determined from the solar wind
measurements \citep{2002JGRAHu}. The setup of such a local
reconstruction frame $R\phi Z$ or $RSZ$ in Cartesian is shown in
Figure~\ref{fig:RSZ}, right panel, where the latter $R$  in $RSZ$ is fixed
corresponding to the radial distance from $Z$ axis at the point of
exit of the spacecraft from the torus. The spacecraft path along
$-\mathbf{r}_{sc}$ with spatially distributed data points (via the
usual transformation of a constant HT frame speed, $V_{HT}$) is
rotated onto the light-shaded $RZ$ plane, where the spacecraft
path is projected approximately onto the   dimension $r$ of
$\theta\approx \theta_0=Const$ in the alternative
$(r,\theta,-\phi)$ coordinate. A cross section is obtained by
solving the GS equation~(\ref{eq:GSrth}) on the light-shaded
plane, utilizing spacecraft measurements along $r$ at
$\theta\approx\theta_0$, as spatial initial values, similar to the
straight-cylinder case. However the distinction here is that due
to the toroidal geometry, the projection onto the cross-sectional
plane is not as straightforward as before. A rotation, rather than
a simple direct projection, has to be performed. For brevity and
completeness, we describe the details of determining both the
origins $O$ and the corresponding radial distance array $R$ along
the spacecraft path in the Appendix~\ref{app:R}, and also describe
the details of the numerical GS solver in the local spherical
polar coordinate $(r,\theta)$ in the Appendix~\ref{app:solver}.

In what follows, we describe, in  details, the two-step
procedures  for determining the $Z$ axis orientation and its
location in terms of its intersection with the $r_{sc}t$ plane,
$O'$, which, in turn, yields the size of the major radius of the
torus. As before, this is implemented in a trial-and-error process
with the location of $O'$ distributed over a finite-size grid on
the $r_{sc}t$ plane, each denoted by the pair $(\rho,\Theta)$, as
shown in Figure~\ref{fig:RSZ}, left panel. Then all possible $Z$
axis orientations are enumerated at each $O'$ location. The
current implementation is such that $\rho\in[0,1)$ AU of a uniform
grid size 0.05 AU, and $\Theta\in[0,2\pi)$ of a uniform grid size
$\pi/20$ for a spacecraft located at a radial distance 1~AU from
the Sun, but excluding $\Theta=0$ and $\pi$ (see
Section~\ref{subsec:de}). At each location $O'$, a trial $Z$ axis
of a unit vector is varied with its arrow tip running over a
hemisphere of unit radius. Associated with each arrow tip, a
residue (see equation~\ref{eq:Rf}) is calculated based on the
theoretical consideration of finding a functional $F$ that best
satisfies the requirement of being a single-valued function of
$\Psi$, based on the GS equation~(\ref{eq:GSt}) (omitting plasma
pressure for the time being; in other words, considering low
$\beta$ plasma configuration only).

\begin{enumerate}
\item The first step is to determine the $Z$ axis orientation via a minimization procedure of the  residue defined in
equation~(\ref{eq:Rf}). This is done by a trial-and-error process
as before, but over the finite-size grid on the $r_{sc}t$ plane.
As shown in Figure~\ref{fig:RSZ} (left panel), at each grid point
$(\rho,\Theta)$, the trial unit $Z$ axis is varying over a
hemisphere of unit radius. For each trial $Z$ axis, the local
reconstruction frame is set up as shown in Figure~\ref{fig:RSZ},
right panel, then the usual transformation from time-series data
to spatially distributed data along the spacecraft path is
performed, together with proper projection (rotation in the
present case) to obtain data along the ``projected" spacecraft
path at $\theta\approx \theta_0$ on the light-shaded cross-sectional
plane. Then the flux function along $r$ at $\theta=\theta_0$ is
calculated
\begin{equation}
\Psi_{sc}(r,\theta=\theta_0) = \int_{r(1)}^{r} RB_\theta dr,
\label{eq:Psi0}
\end{equation} implying $\Psi(r(1),\theta_0)=0$.
Conforming to the straight-cylinder case, a  residue $Res$ is
calculated following exactly the same definition as given in
\citet{2004JGRAHu} to quantitatively assess the satisfaction of
the requirement that the functional $F$ be single-valued across
the toroidal flux rope, i.e., quantifying  the deviation between the $F$
values measured along the overlapping inbound (denoted ``1st") and
outbound (``2nd") branch along the spacecraft path, as represented
by circles and stars in Figure~\ref{fig:RSZ} (right panel; see
Figure~\ref{fig:pta003} for an example), respectively:
\begin{equation}
{Res}=
\frac{[\sum_i(F_i^{\mathrm{1st}}-F_i^{\mathrm{2nd}})^2]^{\frac{1}{2}}}{|\Delta
F|}, \label{eq:Rf}
\end{equation} where the index $i$ runs through an abscissa spanning the range of $\Psi$ value of overlapping branches, and  the normalization factor $\Delta F$
represents the corresponding range of the functional value $F$ over
the two branches. Then the optimal $Z$ axis orientation is chosen
as the direction of minimum $R_f$ within certain error bound,
among the set of locations of $O'$.
\item The second step is to re-run Step~I with the chosen $Z$ axis
orientation and a proper evaluation of $\chi^2$ with measurement
uncertainties  over the $(\rho,\Theta)$ grid. The quantity
$\chi^2$ is defined according to \citet{2002nrca.book.....P} to
evaluate the {\em goodness-of-fit} between the measured magnetic
field $\mathbf{B}$ and the GS model output $\mathbf{b}$ along the
spacecraft path, with given uncertainties (e.g., those available from
NASA CDAWeb for Wind spacecraft measurements) $\sigma$:
\begin{equation}
\chi^2=\sum_{\nu=X,Y,Z}\sum_{i=1}^N\frac{(b_{\nu i} - B_{\nu
i})^2}{\sigma_{\nu i}^2}. \label{eq:chi2}\end{equation} Often a
reduced $\chi^2$ value is obtained by dividing the above by the
degree-of-freedom ($\tt{dof}$) of the system. Since in producing
$\mathbf{b}$, a polynomial fit of order $m$ (usually 2 or 3) is
performed  for $F(r,\theta=\theta_0)$ versus
$\Psi_{sc}(r,\theta=\theta_0)$, it follows ${\tt{dof}}=3N-m-1$. Then
the usage of this step is to yield a unique pair
$(\rho_{min},\Theta_{min})$ at which the corresponding reduced
$\chi^2$ value reaches minimum, $\chi^2_{min}$, for the $Z$ axis
orientation determined in Step~I. In addition, a quantity $Q$
indicating the probability of  a value greater than the
specific $\chi^2$ value is also obtained
\begin{equation} Q= 1 - \tt{chi2cdf}(\chi^2, \tt{dof}), \label{eq:Q}\end{equation}
where the function $\tt{chi2cdf}$ is the cumulative distribution
function of $\chi^2$ as implemented, for example, in Matlab. The
associated uncertainty bounds can  be assessed for various output
based on the standard $\chi^2$ statistics
\citep{2002nrca.book.....P}.
\end{enumerate}
These are the two essential steps we develop to carry out the GS
reconstruction in a general toroidal geometry that have been
implemented in Matlab (the code is included in the auxiliary
material accompanying this article). The additional details, such
as the construction of the reference frame $RSZ$ illustrated in
Figure~\ref{fig:RSZ}, and the final step of computing the
numerical solution of $\Psi$ over an annular region on the cross
section of the torus, utilizing equation~(\ref{eq:GSrth}), are
given in the Appendices. In short, the coordinate system $RSZ$ as
illustrated in Figure~\ref{fig:RSZ}, right panel, is used to
obtain the projection of $\mathbf{r}_{sc}$ onto $r$ at
$\theta\approx\theta_0$. Afterwards, the working coordinate system
is switched to $(r,\theta)$ in which both Steps~I and II are
carried out.

We also caution that the toroidal GS reconstruction we present
here applies to the situation of a spacecraft path exiting into
the ``hole" of the torus, but not to a situation of a
spacecraft path  submerged  within the torus, i.e., not crossing
through into the ``hole". This particular case would yield a
``projected" spacecraft path departing significantly from  a
single coordinate line $\theta=\theta_0$ which renders a numerical
solution to the GS equation impossible. We discuss in
Appendix~\ref{app:hodo} in more detail what the indications are in
terms of the magnetic hodograms from in-situ spacecraft
measurements for such paths.

\subsection{A Degenerated Case} \label{subsec:de}
Before we proceed to present benchmark studies of GS
reconstruction of general toroidal configurations, following the
aforementioned steps, we  single out one special case that needs
special treatment. This is the case of the rotation axis $Z$ being
along $\mathbf{r}_{sc}$, i.e., for $\Theta=0,\pi$, in
Figure~\ref{fig:RSZ}. In this case, a degeneracy occurs such that
the residue remains the same for all trial axis lying on the plane
spanned by $\mathbf{r}_{sc}$ and the true $Z$ axis.

\begin{figure}
 \centerline{\includegraphics[width=0.45\textwidth,clip=]{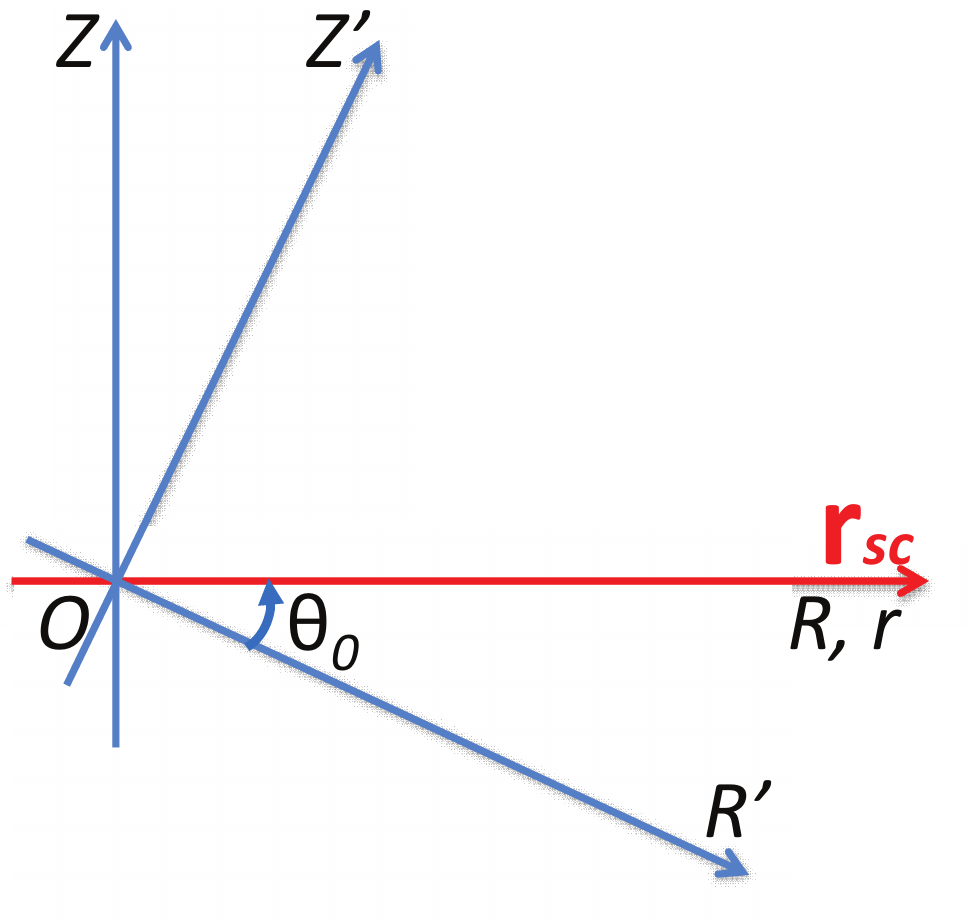}
  \includegraphics[width=0.5\textwidth,clip=]{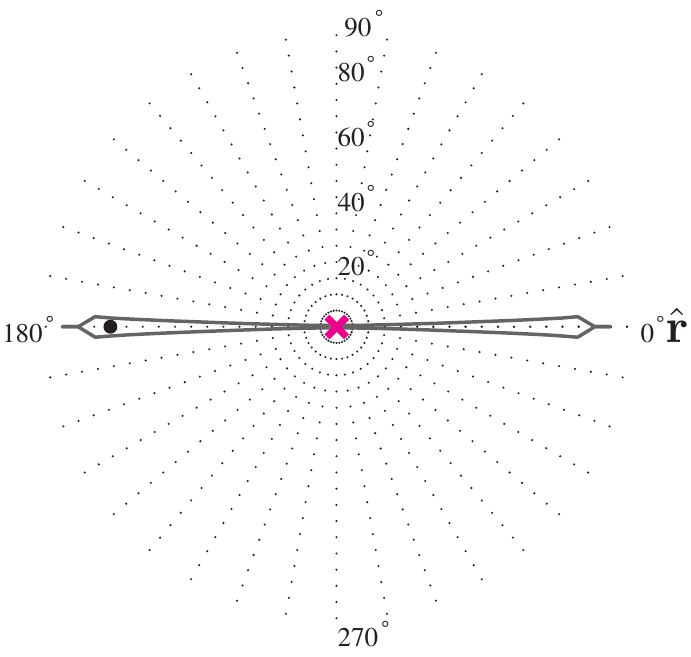}}
 \caption{Left panel: the plane spanned by
 $\mathbf{r}_{sc}$ and $Z$, and another set of trial axes $Z'$ and $R'$ rotated away on the same plane.
 Right panel: the corresponding residue map \citep{2002JGRAHu}. Each point in the background represents an arrow tip of a unit vector (trial $Z$ axis) on the hemisphere centered around
 direction $n$ (center point). The longitudinal separation is 10$^\circ$, while the latitudinal increment is 5$^\circ$. The direction along 0$^\circ$
 longitude is along the direction of $\mathbf{r}_{sc}$, denoted $\hat\mathbf{r}$. The 90$^\circ$ longitude is along $t$.
 The solid contours are drawn at levels $\min(Res)$ and $\min(Res)+1$. The thick dot marks the direction along which the absolute
 minimum value of $Res$ is reached, while the cross marks the true $Z$ axis direction in this case. }\label{fig:Zd}
 \end{figure}
Such degeneration can be understood as follows. As illustrated in
Figure~\ref{fig:Zd}, left panel, all calculations are simply carried out in
the plane spanned by $\mathbf{r}_{sc}$ and the true rotation axis.
Then consider two cases: one with a $Z$ being perpendicular to
$\mathbf{r}_{sc}$ and the other with $Z'$ arbitrarily chosen as
shown. For the former, the composite functional is $F=RB_{\phi}$,
and the flux function along the spacecraft path is, according to
equation~(\ref{eq:Psi0}), $\Psi_{sc}=\int_{r(1)}^r RB_Z dr$
($r\equiv R$). For the other case, correspondingly, $F'=R'B_\phi=R
B_\phi\cos\theta_0$, and
$$\Psi'_{sc}=\int_{r(1)}^r R'B_\theta dr=\int_{r(1)}^r R B_Z\cos\theta_0 dr.$$
Therefore, it  results $F/F'=\Psi_{sc}/\Psi'_{sc}$, given that
 the field components $B_\phi$ and $B_\theta=B_Z$ remain the same,  and the above integrals are always evaluated along
 $\mathbf{r}_{sc}=r \hat\mathbf{r}$, in both
 cases.
Since both functional values $F$ and $\Psi_{sc}$ change with the
$Z$ axis orientation in the same proportion, the  residue of
$F(\Psi)$ remains unchanged for any trial $Z$ axis in the plane.
An example of such a residue map is shown in Figure~\ref{fig:Zd},
right panel, where the residue remains the same for any $Z$ axis
that is lying on the plane spanned by the true $Z$ axis (along
$n$) and $\mathbf{r}_{sc}$. Note that this behavior does not
change with added noise since the derivation shown above still
applies no matter whether or not noise is added.

In practice, such a degenerated case presented above in
Section~\ref{subsec:de} can either be run separately or simply
excluded, considering that this case may be encompassed by the
uncertainty regions of the most general cases discussed in
Section~\ref{subsec:general}, as to be illustrated below in
benchmark studies. Alternatively, since such degeneracy only
affects Step~I the most, one may still include these grid points along
$\Theta=0$ and $\pi$ in Step~II, once an optimal $Z$ axis has been
determined.

\section{Benchmark Studies}\label{sec:bench}
The benchmark studies of the reconstruction procedures are carried
out against a set of analytic solutions to the GS
equation~(\ref{eq:GSt}) that has been well studied in fusion
plasmas. In particular, such solutions were given by
\citet{freidberg87} for 2D toroidal configurations (for additional
details and variations, see \citet{2010PhPlC}). We provide below
such analytic formulas in terms of the flux function as a function
of space in the $(R,\phi,Z)$ coordinates and associated parameters
defining the overall geometry that forms the basis of analysis in
this Section.

From \citet{freidberg87} (Chapter~6, pp.~162-167), an exact
solution to GS equation~(\ref{eq:GSt}) exists for a special known
functional form of the right-hand side, i.e., $FF'=A=Const$ and
$-\mu_0 p'=C=Const$, and can be written
\begin{equation}
\Psi=\frac{C\gamma}{8}[(R^2-R_a^2)^2-R_b^4]+\frac{C}{2}[(1-\gamma)R^2]Z^2-\frac{1}{2}AZ^2,\label{eq:PsiRZ}
\end{equation}
with $R_a^2=r_0^2(1+\epsilon^2)$ and $R_b^2=2r_0^2\epsilon$ where
the ratio between the minor and major radii of the torus is
$\epsilon=a/r_0$. {\textbf{The geometry of the cross section of
the torus is completely determined by the parameters $r_0$ and
$a$, which define the center $R=r_0$ and the boundary $R=r_0\pm a$
of the cross section at $Z=0$. The other constant
$\gamma=\frac{\kappa^2}{1+\kappa^2}$ is related to the plasma
``elongation", $\kappa$, in confinement devices, defined as the
ratio between the area of the plasma cross section and $\pi a^2$
\citep{freidberg87}.}} Now we start to deviate from
\citet{freidberg87} referenced above since our purpose is to
utilize the solution provided by equation~(\ref{eq:PsiRZ}), but
not to follow the subsequent analysis of the properties of such a
solution.

By normalizing both spatial dimensions by $r_0$, i.e., $R=xr_0$
and $Z=yr_0$, we obtain
\begin{equation}
\Psi=\Psi_0\left[x^2-1+\frac{1-\gamma}{\gamma}\frac{1+\epsilon^2}{\epsilon^2}
(1+\frac{2\epsilon}{1+\epsilon^2}x)y^2-\frac{1}{2}
\frac{A}{\Psi_0/r_0^2}y^2\right].\label{eq:Psixy}
\end{equation}
We choose the following parameter values to obtain solutions that
yield reasonable geometric dimensions and magnetic field magnitude
consistent with in-situ MC observations at 1 AU: $\epsilon=0.1$,
$\gamma=0.8$, and $\Psi_0/r_0^2=1$ nT, $A=-40$ or -10 nT. Then the
transverse field components $B_R$ and $B_Z$ (equivalently, $B_r$
and $B_\theta$) are obtained from equation~(\ref{eq:B}). The axial
field $B_\phi$ is determined from $F^2=2A\Psi+B_0^2$, where the
integration constant $B_0$ is arbitrarily chosen.

The time-series data for analysis are obtained by flying a virtue
spacecraft through such a torus along a pre-set path, in the
direction opposite to $\mathbf{r}_{sc}$ (for different
perspectives, see Figures~\ref{fig:RSZ} and \ref{fig:psi332}).
Then the magnetic field vectors $\mathbf{B}$ extracted from the
analytic solution described above along this path in $r_{sc}tn$
coordinate are  further modified by adding normally distributed
noise component-wise  up to certain level characterized by the
quantity $\tt{NL}$:
\begin{equation}
\tilde\mathbf{B}=\mathbf{B} +
\tt{randn()}*NL*\langle|\mathbf{B}|\rangle, \label{eq:dB}
\end{equation}
where the random number generator $\tt{randn()}$ yields numbers from a  normal
distribution of zero mean and unit standard deviation. Therefore
each magnetic field component in the time series for the following
analysis
 carries a constant standard deviation in each case
 $\sigma=\tt{NL}*\langle|\mathbf{B}|\rangle$.

 Note that in the following benchmark studies, we omit the
 pressure gradient in the right-hand side of the GS equation
 completely, although the exact solution we test against does
 include a finite pressure distribution ($C\ne 0$; otherwise the solution is trivial). This is based on the
 consideration that in real applications to mostly low $\beta$
 flux rope structures in the solar wind, the plasma
 pressure is usually less important and carries relatively larger
 measurement uncertainties. So the current GS model outputs for the
 toroidal geometry, i.e., the determination of $Z$ and $r_0$, are primarily based on the magnetic field
 measurements. The measurements of plasma  pressure, of course, will be included in applications to real events.

\subsection{Determination of $Z$ and $r_0$}\label{subsec:Z}

\begin{figure}
 \centerline{\includegraphics[width=1.\textwidth,clip=]{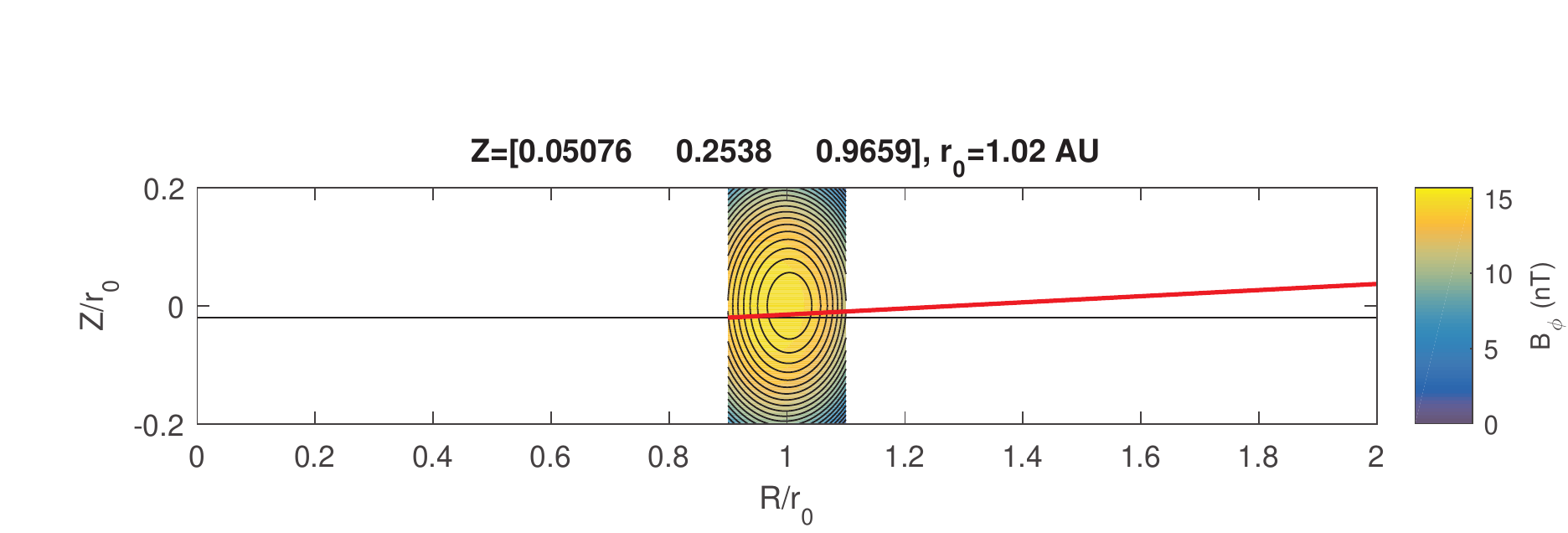}}
 \caption{One analytic solution given by equation~(\ref{eq:Psixy}) with $A=-40$ nT. Contours represent the flux function
 $\Psi$ on the $RZ$ plane. Color indicates the axial component $B_\phi$, with scales given by the colorbar. The main geometrical parameters
 $Z$ in $r_{sc}tn$ and $r_0$ are given on top. The red line denotes the projected spacecraft path $r$ and the horizontal
 black line the $R$ axis intersecting $r$ at $R_0$, corresponding to Figure~\ref{fig:RSZ}, right panel. }\label{fig:psi332}
 \end{figure}
In this section, we present one example of  benchmark studies to
show the results of determining the orientation and location of
the rotation axis $Z$, i.e., in turn, the major radius $r_0$,
following the steps outlined in Section~\ref{sec:proc}. Although a
number of additional benchmark studies was carried out, based on
the analytic solution of equation~(\ref{eq:Psixy}) with different
configurations, i.e.,  different virtue spacecraft paths across,
and different noise levels, it is not possible to study and
present all cases in an exhaustive manner. Therefore we choose to
present one case and are providing the computer codes in Matlab to
encourage the interested users to repeat or generate new results,
and to follow up with their own studies.

\begin{figure}
 \centerline{\includegraphics[width=0.5\textwidth,clip=]{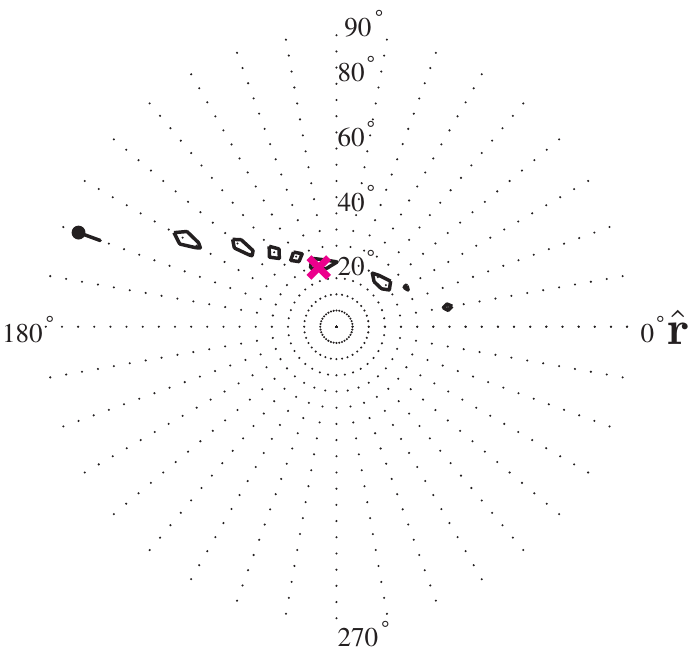}
 \includegraphics[width=0.5\textwidth,clip=]{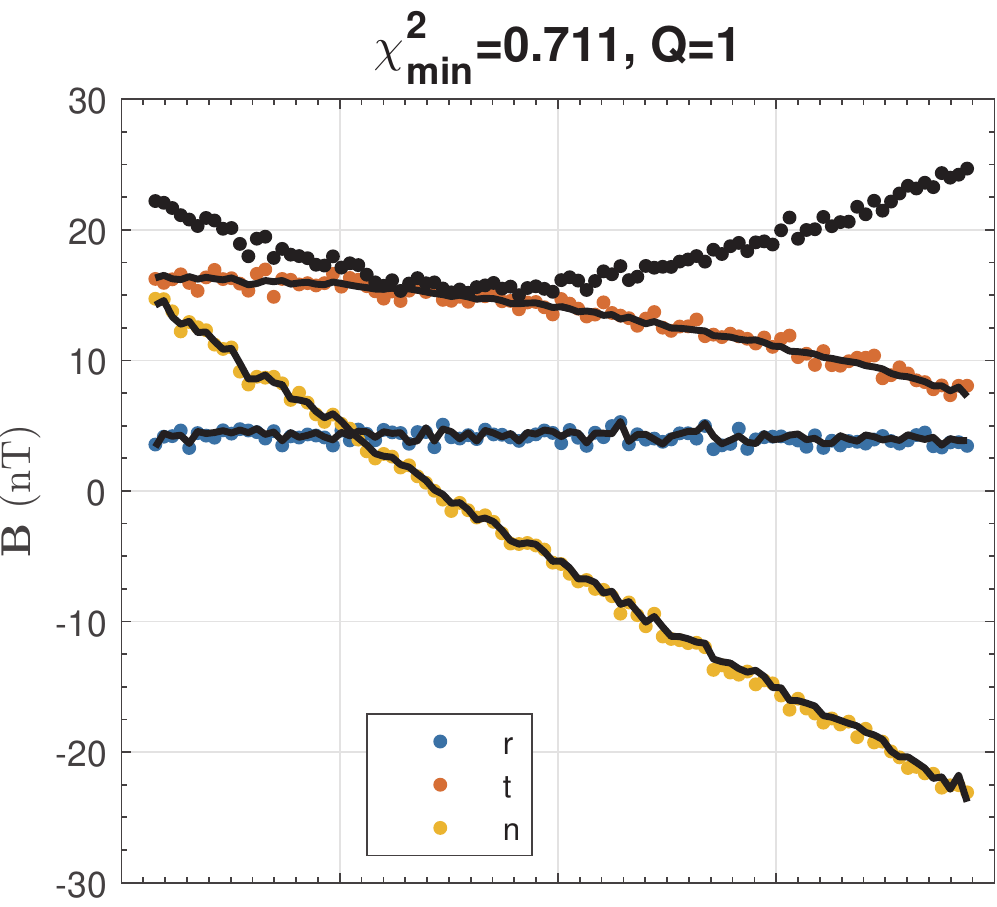}}
 \caption{Left panel: The residue map at the $O'$ location where the minimum value was
 obtained. Format is the same as Figure~\ref{fig:Zd}, right panel. The thick dot marks the direction along which the absolute
 minimum value of $Res$ was reached, while the cross marks the final $Z$ direction chosen based on the distribution of residues on this residue map.
 Right panel: The magnetic field  data along the spacecraft path (see legend; {\textbf{here $r\equiv r_{sc}$}}) and the corresponding GS model outputs (black curves)
 The resulting minimum reduced $\chi^2$ and the associated $Q$ values are given on top. }\label{fig:Zr0}
 \end{figure}

Figure~\ref{fig:psi332} shows the overall configuration of this
benchmark case in the $RZ$ plane, on which the exact solution is
shown within the rectangular domain. The ``projected" spacecraft
path is along the red line of an approximately constant
$\theta\approx\theta_0=3.0^\circ$ formed with the horizontal line
intersecting the cross section at $R=R_0$ (see also
Figure~\ref{fig:RSZ}). The exact $Z$ axis orientation and major
radius of the torus are noted in the title of the figure. The
synthetic time-series data for analysis are obtained along
$-\mathbf{r}_{sc}$ from the analytic solution shown with
additional noise according to equation~(\ref{eq:dB}) for
$\tt{NL}=0.025$ in this case. The resulting time series are shown
in Figure~\ref{fig:Zr0}, right panel, together with the GS model
output to be further discussed.

 \begin{figure}
 \centerline{\includegraphics[width=1.\textwidth,clip=]{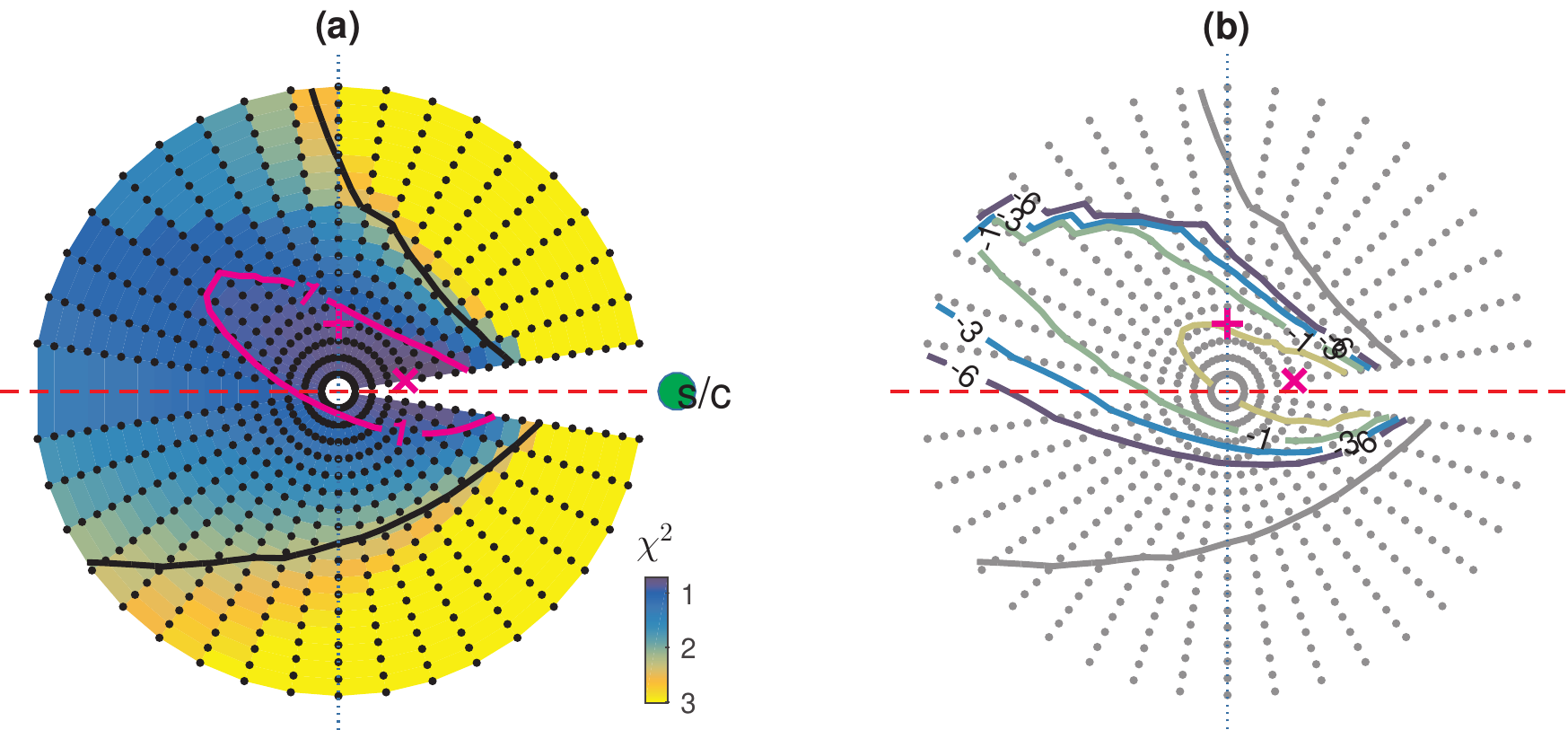}}
 \caption{(a) The distribution of reduced $\chi^2$ value, as indicated by the colorbar  on the $r_{sc}t$ plane. The background
 dots are the grid points in $(\rho,\Theta)$. The contours are of levels 1, and 1+$\sqrt{2}$, respectively. (b) The corresponding contour plot of $\log_{10}Q$ as labeled.
 The innermost contour is of level $Q=0.9$, and the outermost one is the same as the black one in (a).   The plus and cross signs
 mark the true $Z$ axis location and that of $\chi^2_{min}$, respectively.}\label{fig:chi2}
 \end{figure}
We carried out the analysis following the two steps delineated in
Section~\ref{sec:proc} for the most general case, i.e., $Z$ not
along $\mathbf{r}_{sc}$ and not parallel to $n$ either, in this
case. From Step~I, we calculated the residue at each
$(\rho,\Theta)$ grid point based on equation~(\ref{eq:Rf}) and
found the minimum value, $\min(Res)=0.37$. The corresponding
residue map at this particular location where the minimum value
was obtained is shown in Figure~\ref{fig:Zr0}, left panel. The
distribution of residues on this residue map exhibits multiple
local minima, in the form of a string of ``islands", each of value
$\min(Res)+1$. Sometimes, they often merge and form one elongated
shape enclosing a number of grid points. The general rule-of-thumb
based on our experiments and experience is that the optimal $Z$
axis orientation should always be chosen near the middle of either
one large single contour or one single ``island" located near the
middle of the group, as in the present case. Such an axis is
chosen, usually through an interactive, manual process, as marked
by the cross symbol, which is $[-0.09366,   0.3134, 0.9450]$ in
$r_{sc}tn$ coordinate.

With this chosen $Z$ axis, we subsequently carried out Step~II. The
results of the reduced $\chi^2$ distribution and the corresponding
$Q$ values are shown in Figure~\ref{fig:chi2}a and b, respectively.
Equation~(\ref{eq:chi2}) can be used to evaluate the reduced
$\chi^2$ values by replacing the variables $\mathbf{B}$ and
$\mathbf{b}$ by the ones normalized by $\sqrt{\tt{dof}}$. As
stated in \citet{2002nrca.book.....P}, such defined reduced
$\chi^2$ values tend to a  distribution of mean 1 and standard
deviation $\sqrt{2/\tt{dof}}$ (maximum $\sqrt{2}$). A value $\sim
1$ indicates a ``moderately good" fit. Correspondingly, the
probability of such a ``good" fit, $Q$, has to be significant,
e.g., $>0.1$. Therefore, in Figure~\ref{fig:chi2}, two contours
of levels 1 and $1+\sqrt{2}$ are shown for the $\chi^2$
distribution, and a number of contours are shown for $\log_{10}
Q$, with the innermost one of value $Q=0.9$. Combined, the
contours of values $\chi^2=1$ and $Q=0.9$ indicate the extent of
uncertainty in the location of $Z$, i.e., the uncertainty in major
radius. Both the exact location and one selected location of $Z$
where $\chi^2$ reaches minimum are enclosed by the innermost
contours. The corresponding major radii for these two locations
are 1.02 AU and 0.80 AU, respectively. The resulting GS model
output $\mathbf{b}$ components (together with $\mathbf{B}$) along
the spacecraft path for the chosen $Z$ axis orientation and
location of minimum $\chi^2_{\min}=0.711$ ($Q=1$),
 are shown in
Figure~\ref{fig:Zr0}, right panel.

 \begin{table}
 \caption{Comparison of the major geometrical parameters for the benchmark case.}\label{tbl:para}
 \begin{tabular}{ccc}
 \hline
Benchmark & $Z$, $[r_{sc},t,n]$ & $r_0$ (AU)  \\
\hline
Exact & [0.05076, 0.2538, 0.9659] & 1.02 \\
GS & [-0.09366,   0.3134,   0.9450] & 0.80  \\
Error & 9$^\circ$ &  22\%  \\
\hline
 \end{tabular}
 \end{table}
 As summarized in Table~\ref{tbl:para}, the two major geometrical
parameters, namely, the rotation axis $Z$ and major radius $r_0$,
were determined through the above procedures and are compared with
the exact values of this benchmark case. The absolute error in the
$Z$ axis orientation is 9$^\circ$ and that in $r_0$ is about 22\%.
The latter can be regarded as an  uncertainty estimate in $r_0$,
since the separation between the exact and selected $Z$ axis
locations spans approximately the half-width of the maximum extent
of the innermost contours in Figure~\ref{fig:chi2}.

\subsection{Accuracy of the GS Solver}\label{subsec:solver}
\begin{figure}
 \centerline{\includegraphics[width=0.5\textwidth,clip=]{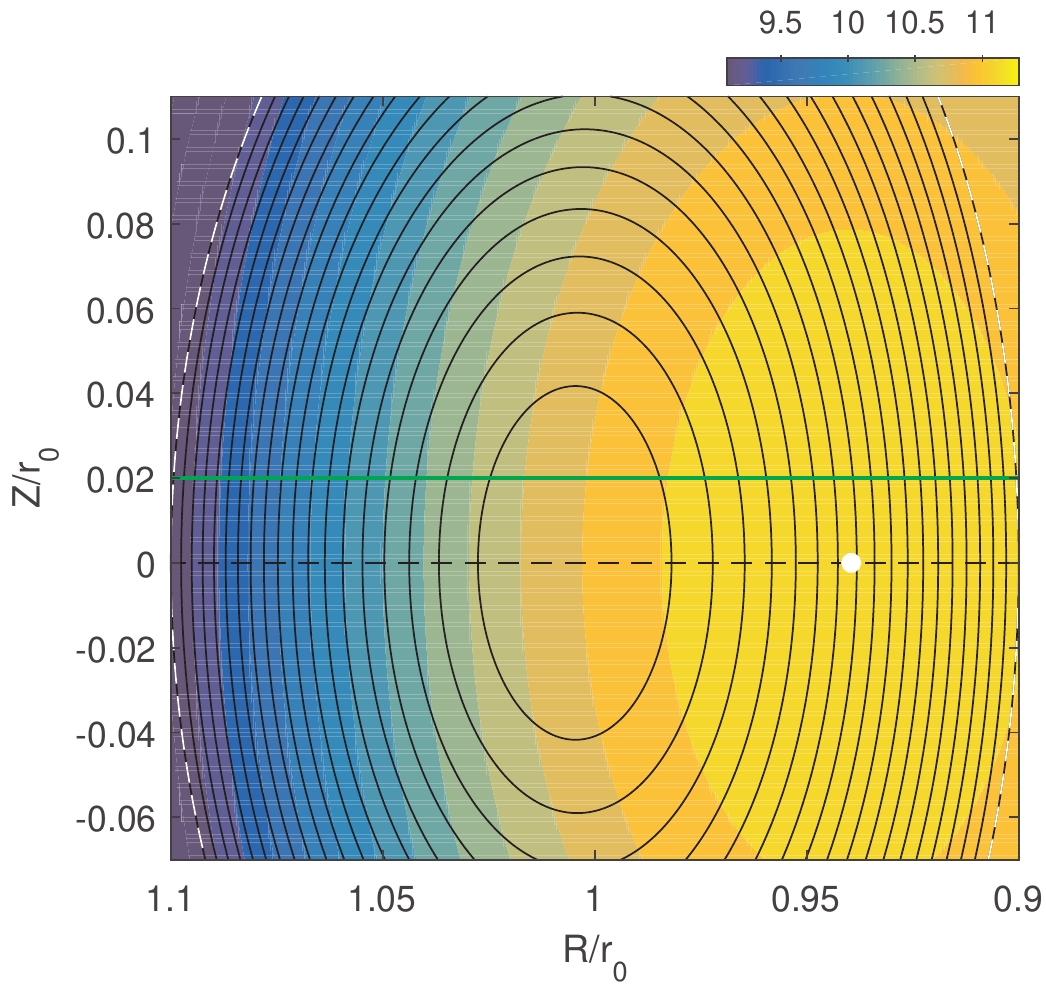}
 \includegraphics[width=0.5\textwidth,clip=]{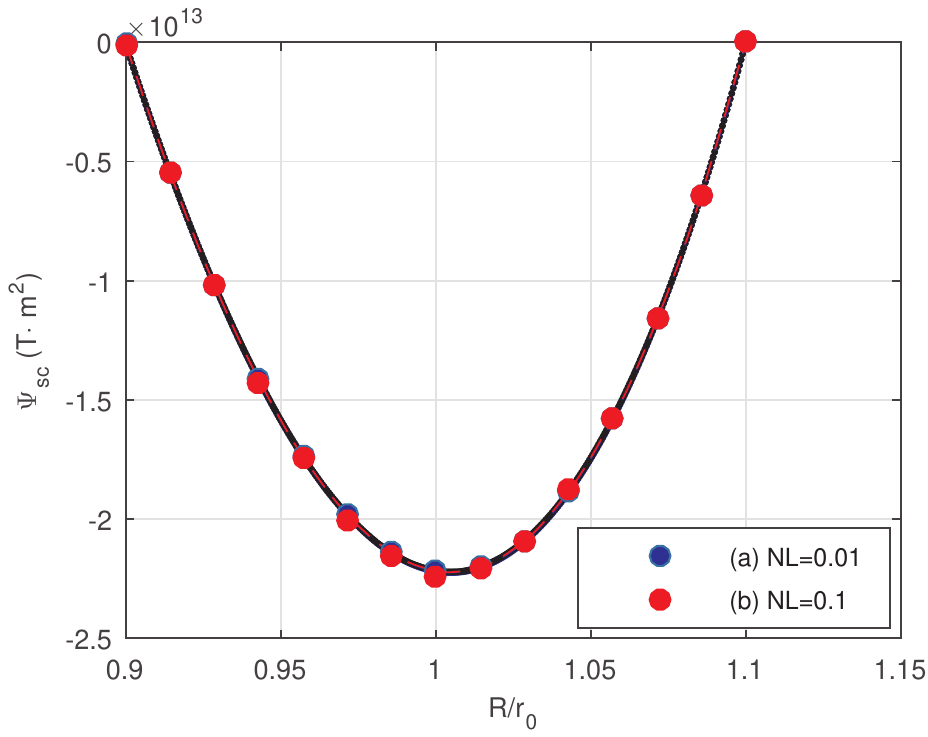}}
 \caption{Left panel: One benchmark case in terms of the analytic solution given on the $RZ$ plane ($r_0=1$ AU). Contours represent the flux
 function $\Psi$, and colors the $B_\phi$ component, as indicated by the colorbar. The dashed contour is of value 0. The horizontal green line denotes the spacecraft path,
 while the dashed line the symmetry line of the solution. White dot marks the location of maximum $B_\phi$. Right panel: the numerically
 calculated flux function values along the spacecraft paths  for the two cases with different noise levels. }\label{fig:psisc}
 \end{figure}
We present, separately in this section, the benchmark studies on
the accuracy of the numerical GS solver with details given in the
Appendix~\ref{app:solver}. The purpose is to test the
implementation  of the solver in the code, and to assess its
performance in terms of error estimates under idealized condition
of an exact set of $Z$ axis and $r_0$, independently  from
Section~\ref{subsec:Z}.

\begin{figure}
 \centerline{\includegraphics[width=0.5\textwidth,clip=]{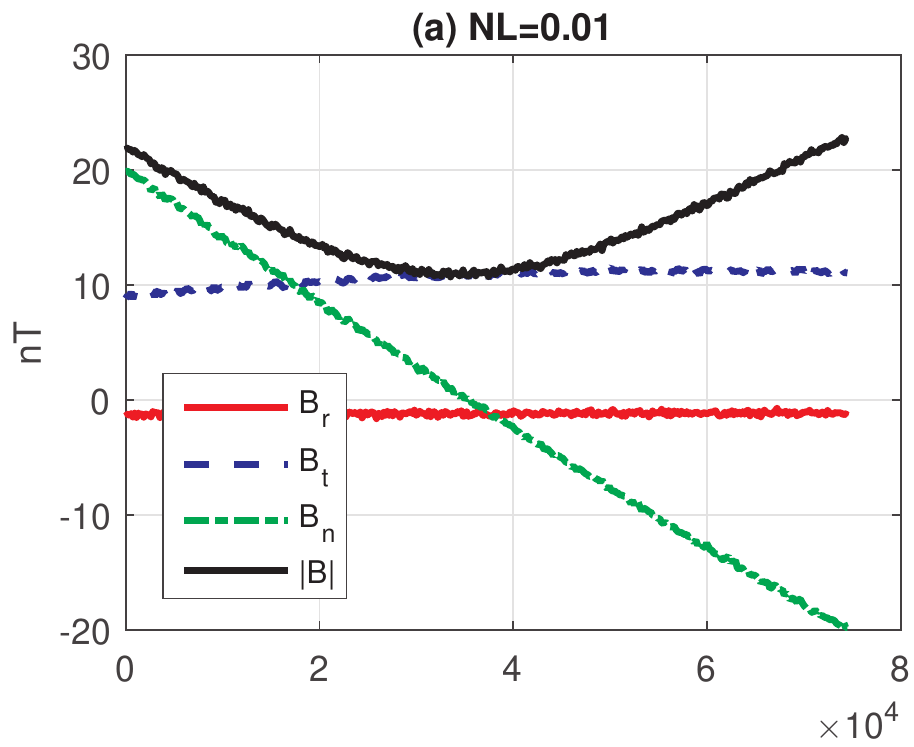}
 \includegraphics[width=0.5\textwidth,clip=]{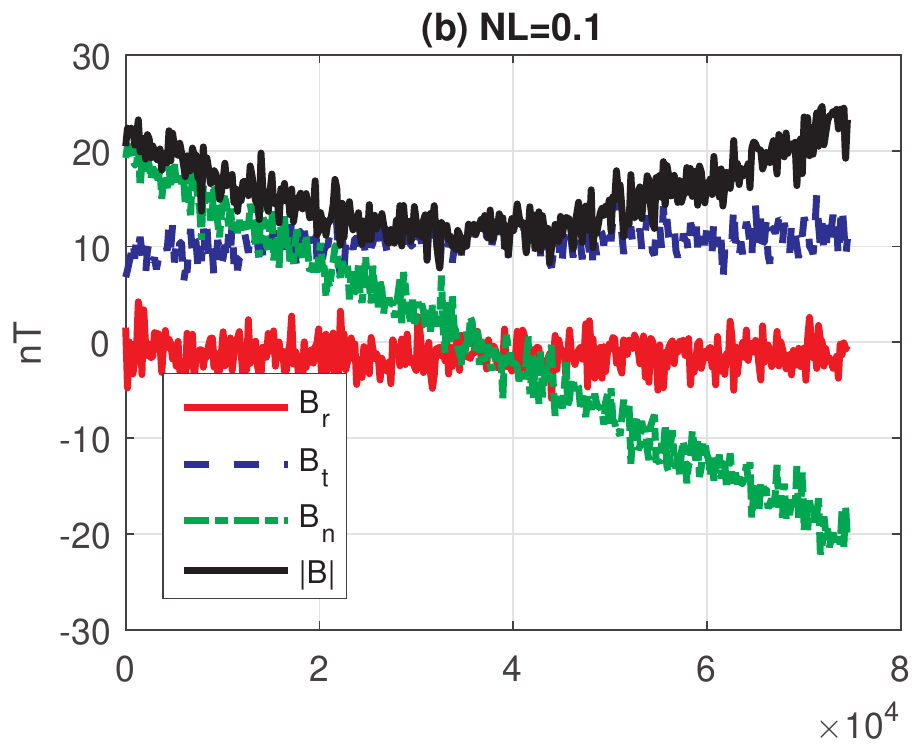}}
 \caption{The magnetic field components and magnitude along the spacecraft path for Case (a) and (b), respectively. Here $r\equiv r_{sc}$.}\label{fig:B003}
 \end{figure}
 Two cases of two different $\tt{NL}$ values are considered, for a geometry of the spacecraft path parallel to
 $R$, i.e., $\theta_0=0$, so that a direct point-by-point
 comparison between the exact  and numerical GS solutions can be
 made with minimal interpolation effect. Such an exact solution is shown in Figure~\ref{fig:psisc} (left panel) where the
 solution is given on the grid in $RZ$ coordinate, while the right panel shows the corresponding numerically calculated
 flux function values along the spacecraft path for the two cases indicated by the legend.  The time series for the
 two cases of different levels of noise added to the exact
 solution are shown in Figure~\ref{fig:B003} for (a) $\tt{NL}=0.01$, and
 (b) $\tt{NL}=0.1$, respectively. Case (b) is used as an extreme
 example to illustrate the effect of noise (see additional results
 below). We observe that a real event in terms of derived quantities  is close to case (a) or
 somewhere in-between case (a) and (b).

 \begin{figure}
 \centerline{\includegraphics[width=0.5\textwidth,clip=]{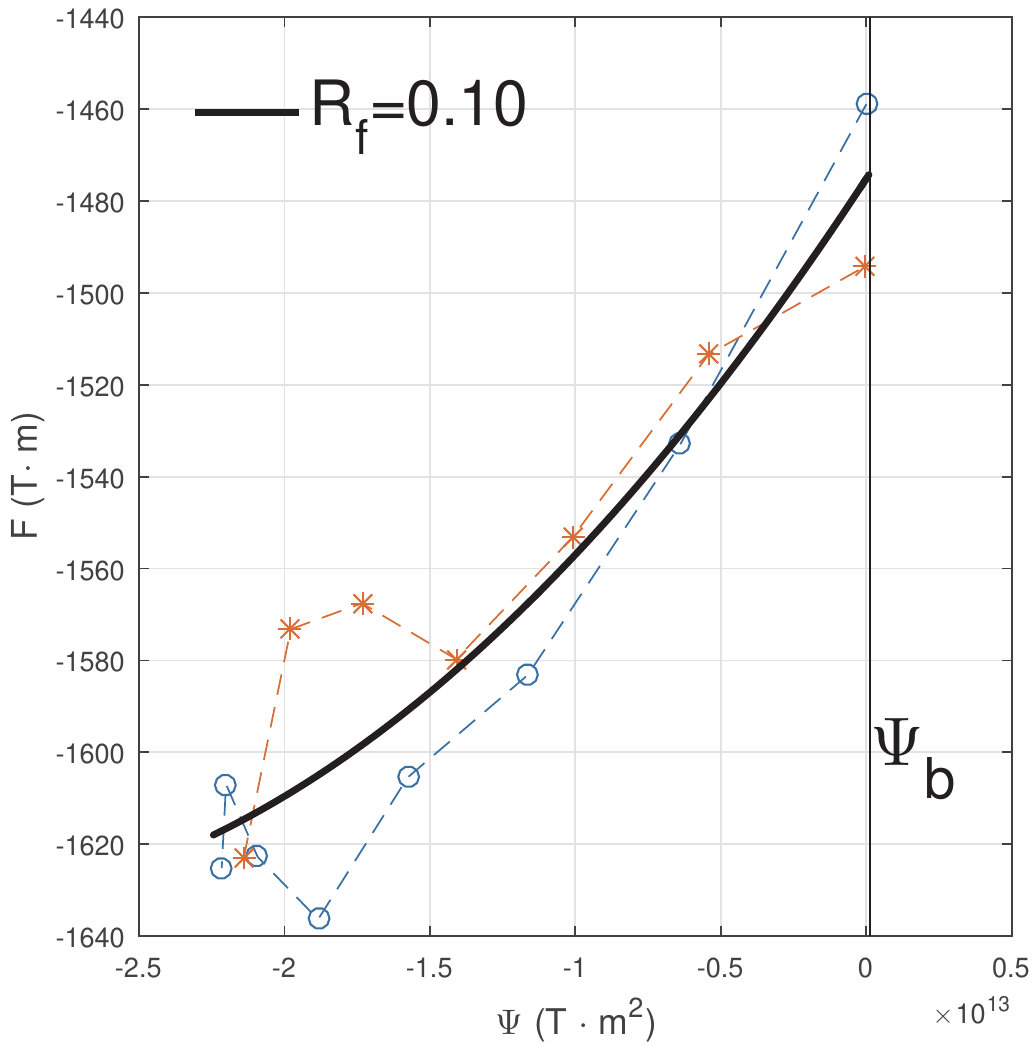}
 \includegraphics[width=0.5\textwidth,clip=]{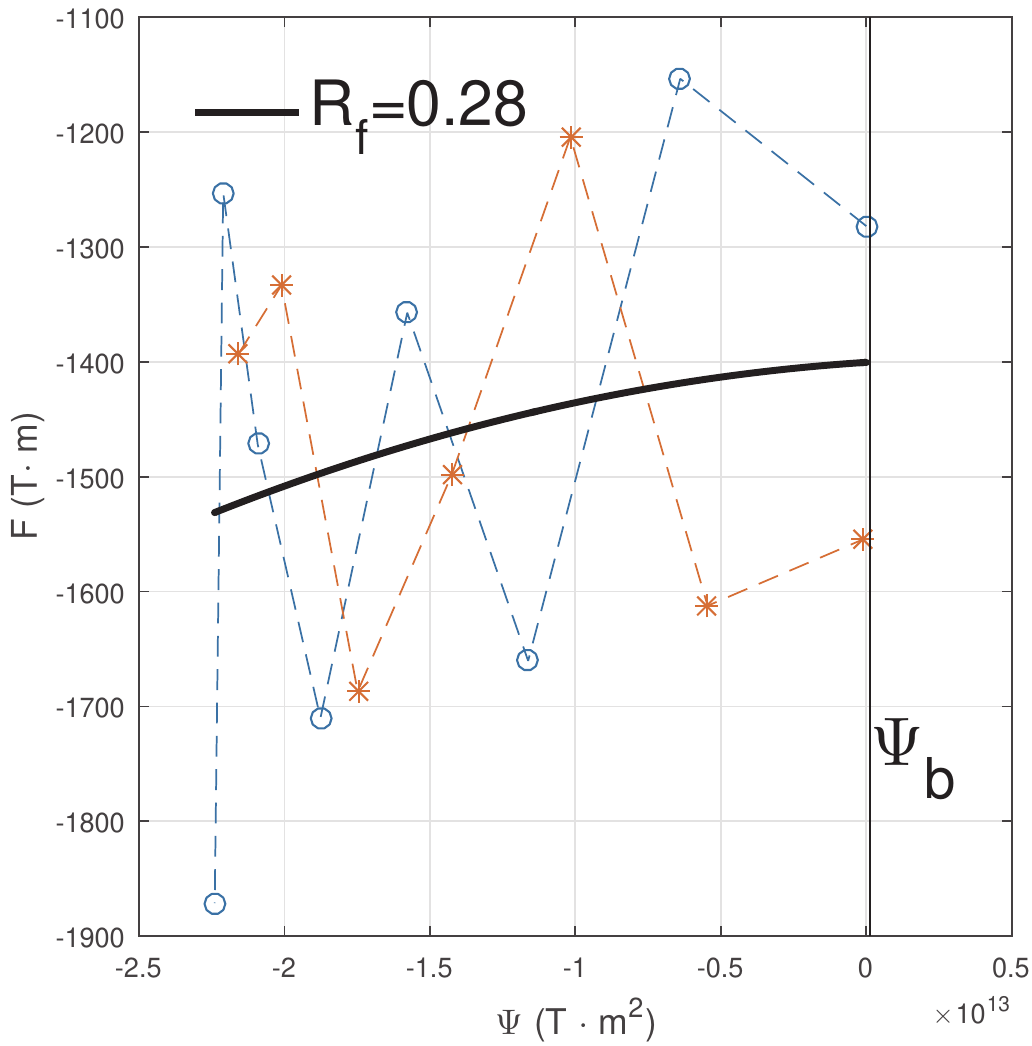}}
         \vspace{-0.50\textwidth}   
     \centerline{ \bf     
      \hspace{0.05 \textwidth} {(a)}
      \hspace{0.45\textwidth}  {(b)}
         \hfill}
     \vspace{0.50\textwidth}    
\caption{The corresponding measured $F$ versus $\Psi$ data points
along the spacecraft path, and the 2nd-order polynomial fitting
 $F(\Psi)$ (black curve) for Case (a) and (b), respectively. A fitting residue $R_f$ and a boundary $\Psi=\Psi_b$ are also marked \citep{2004JGRAHu}. }\label{fig:pta003}
 \end{figure}This is demonstrated by the corresponding $F(\Psi)$ plots and the
 fitting residues $R_f$ \citep{2004JGRAHu} in Figure~\ref{fig:pta003} along the spacecraft path. Case (a) resembles
 what one gets from real data with a typical and relatively small
 fitting residue that is considered acceptable (usually when $R_f<0.20$), indicating
 reasonable satisfaction of the requirement that the functional
 $F(\Psi)$ be single-valued. On the other hand, in case (b), the
 data scattering is large and the fitting residue exceeds 0.20,
 indicating that the satisfaction of $F(\Psi)$ being
 single-valued is questionable. The fitting polynomials are of 2nd
 order in these cases, while the 1st-order polynomials yield
 similar results. In practice, such reconstruction results for case (b)  with
 this  metric value would have been rejected.

\begin{figure}
 \centerline{\includegraphics[width=0.5\textwidth,clip=]{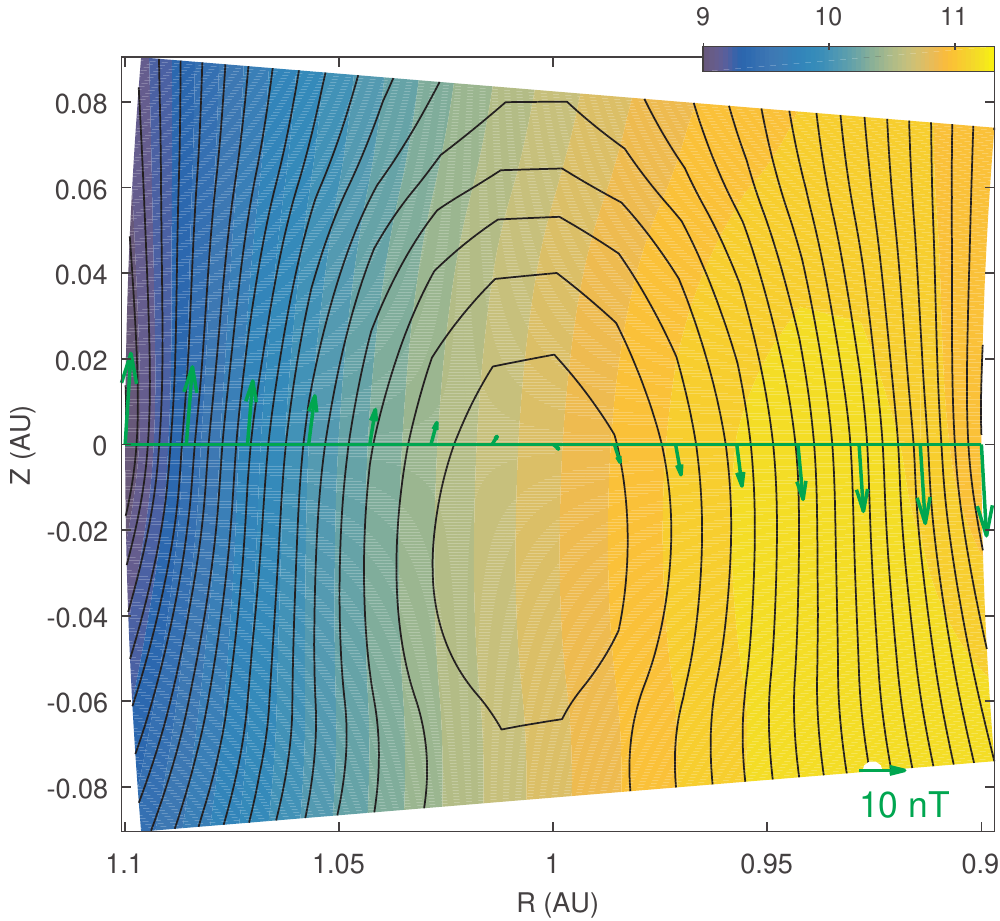}
 \includegraphics[width=0.5\textwidth,clip=]{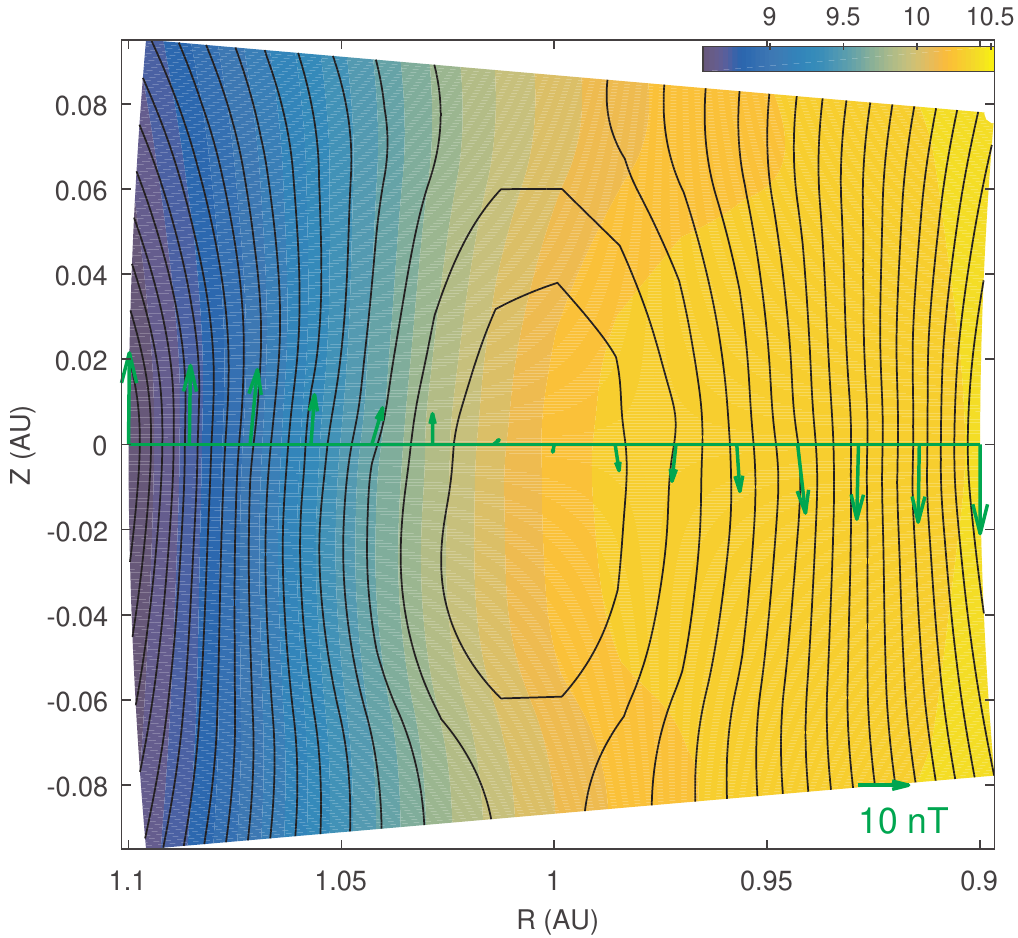}}
        \vspace{-0.42\textwidth}   
     \centerline{ \bf     
      \hspace{0.06 \textwidth} \color{white}{(a)}
      \hspace{0.44\textwidth}  \color{white}{(b)}
         \hfill}
     \vspace{0.42\textwidth}    
 \caption{The resulting numerical solution obtained by the toroidal GS solver for Case (a) and (b), respectively. The format is the same
 as Figure~\ref{fig:psisc}, left panel. Additionally, the green arrows represent the measured transverse magnetic field components along
 the spacecraft path. }\label{fig:map003}
 \end{figure}
\begin{figure}
 \centerline{\includegraphics[width=0.5\textwidth,clip=]{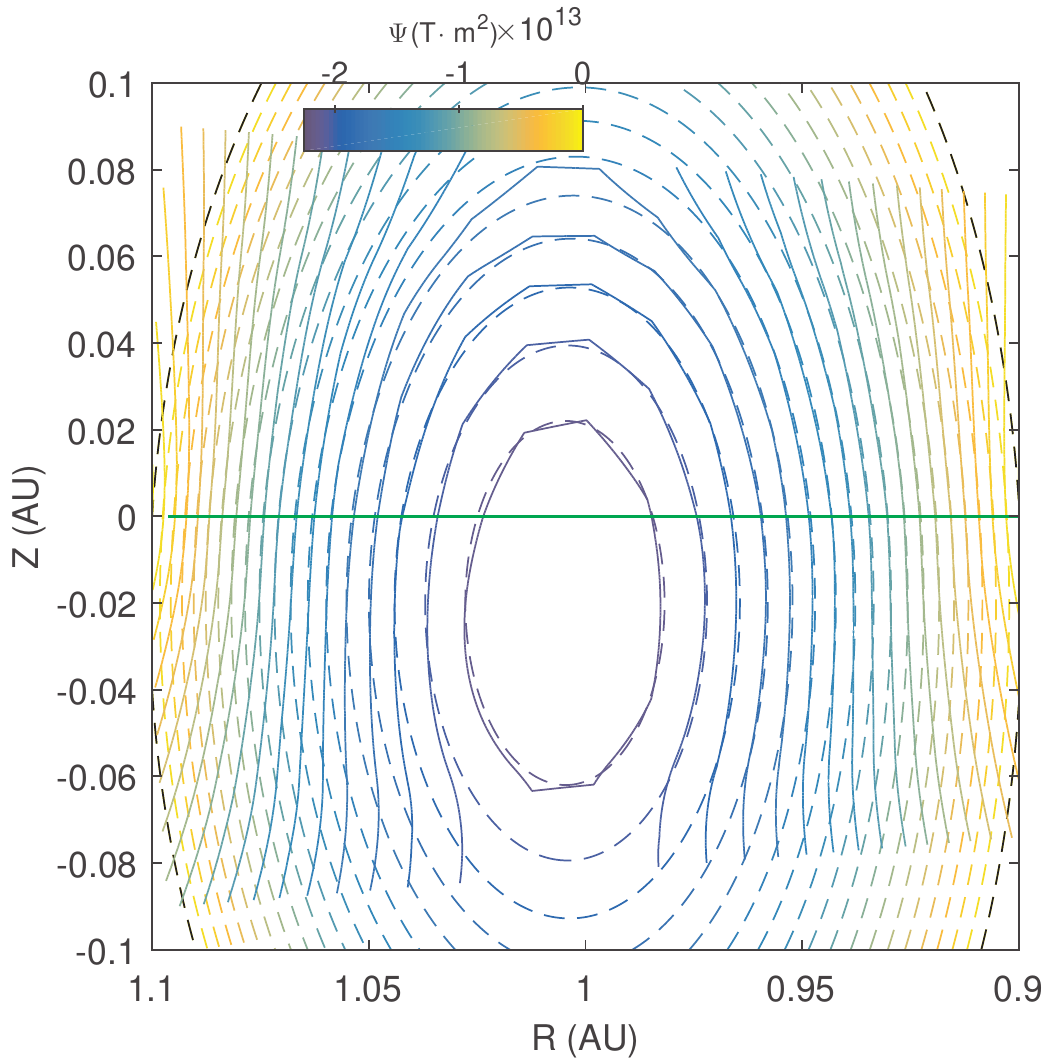}
 \includegraphics[width=0.5\textwidth,clip=]{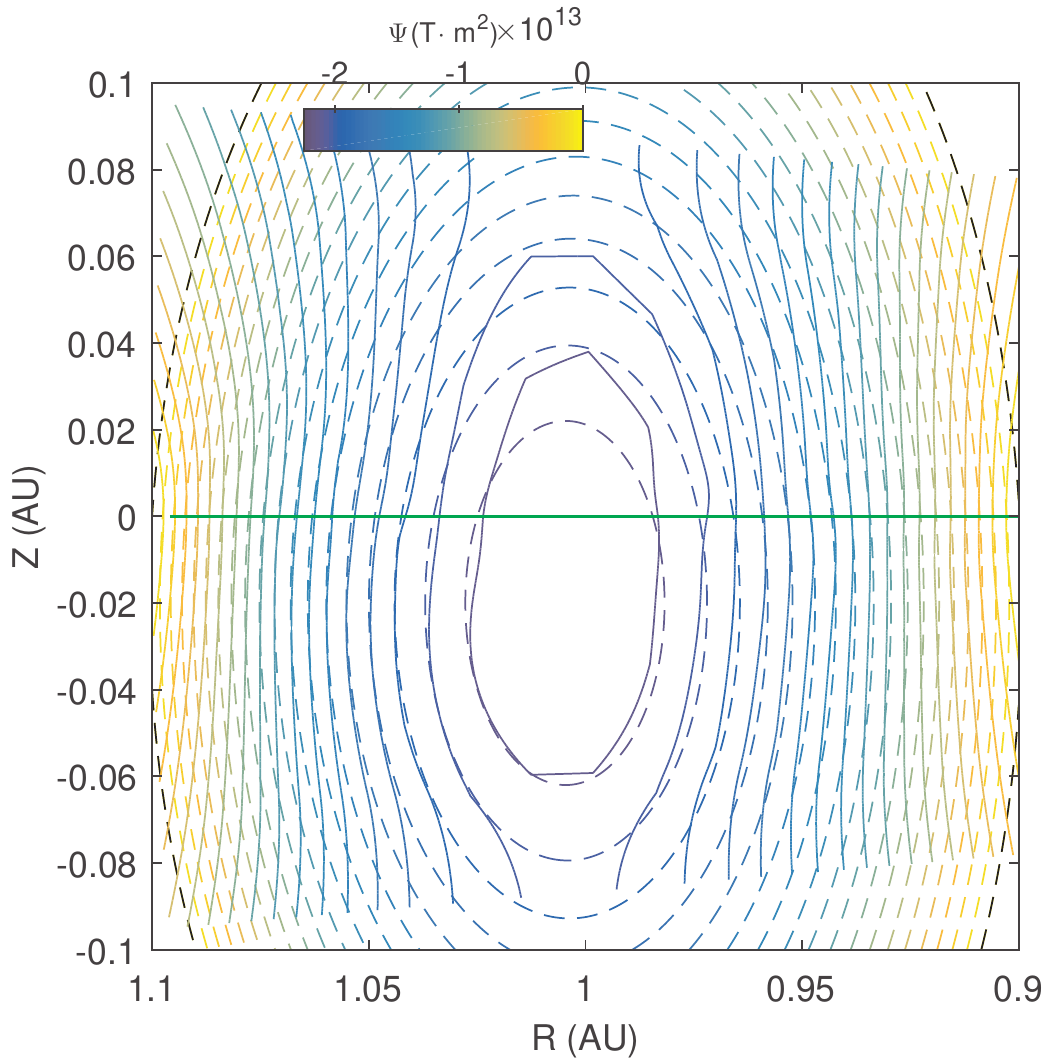}}
         \vspace{-0.51\textwidth}   
     \centerline{ \bf     
      \hspace{0.05 \textwidth} {(a)}
      \hspace{0.45\textwidth}  {(b)}
         \hfill}
     \vspace{0.51\textwidth}
      \caption{The overplotted contours of the exact (dashed lines) and the numerical (solid lines) solutions, for Case (a) and
 (b), respectively. Colorbar indicates the range of $\psi$. }\label{fig:psi003}
 \end{figure}

 The numerical GS reconstruction results for the two cases are
 shown in Figure~\ref{fig:map003} (a) and (b), respectively, in the usual format.
 Compared with the exact solution in Figure~\ref{fig:psisc}, there
 are clear distortions due to noise and numerical errors. The deviations seemingly increase with increasing
 noise levels. The maximum axial field is 11.3 nT and 10.5 nT, respectively, the location of which is also different from
 that of the exact solution. {\textbf{The areas of the strongest $B_\phi$ seem to be distorted or shrunk compared with Figure~\ref{fig:psisc} (left panel),
 due to the errors which directly affect the evaluation of $F(\Psi)$ in obtaining $B_\phi$. }}
  To further assess, quantitatively, the numerical
 errors, Figure~\ref{fig:psi003} shows the contour plots of the
 flux function, with both the exact solution $\Psi$ and the
 numerical solution $\psi$, overplotted on the same set of contour
 levels, for both cases. It becomes clearer that case (a) solution
 agrees better with the exact solution than case (b). The range of
 the $\psi$ values, representing the amount of poloidal flux $\Phi_p$, for
 both cases, is well recovered, as indicated by the colorbar. This
 agrees with
  Figure~\ref{fig:psisc}, right panel, where the calculated flux functions along the spacecraft path for both cases, although case (b)
 exhibits slightly larger errors, agree with the exact values
 well. This indicates the effectiveness  of low-pass filtering we
 carry out at the beginning of the analysis in processing the time-series data.

\begin{figure}
 \centerline{\includegraphics[width=0.5\textwidth,clip=]{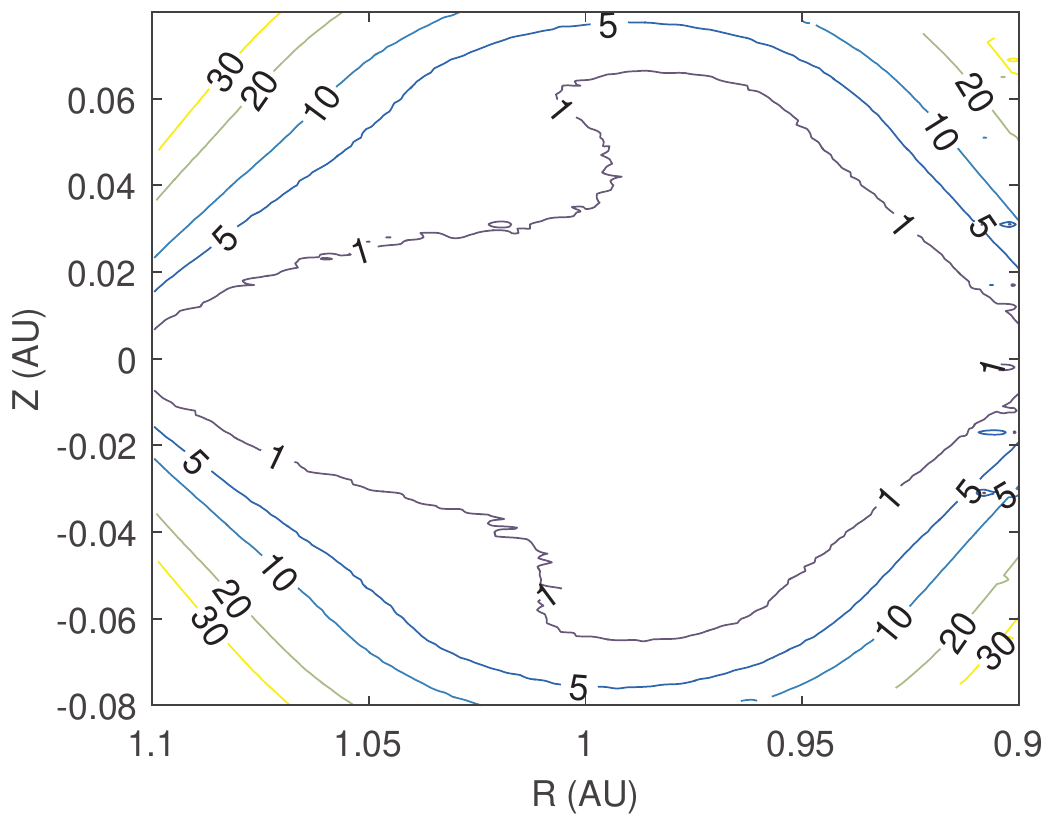}
 \includegraphics[width=0.5\textwidth,clip=]{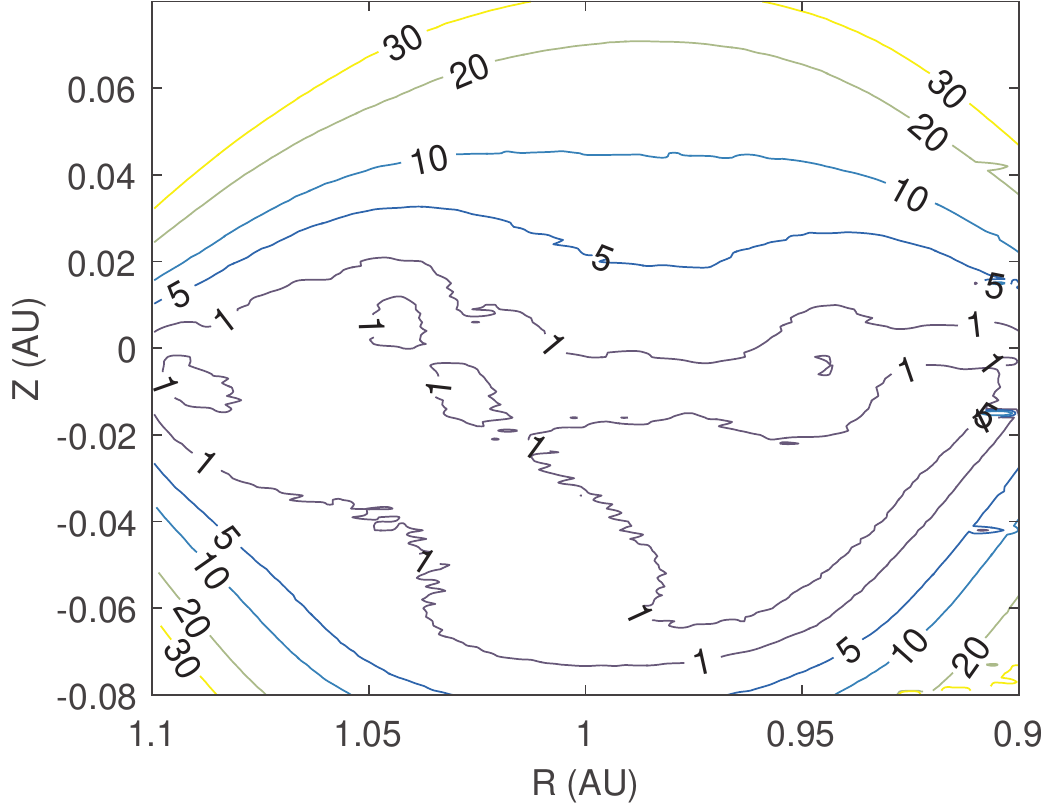}}
          \vspace{-0.42\textwidth}   
     \centerline{ \bf     
      \hspace{0.05 \textwidth} {(a)}
      \hspace{0.45\textwidth}  {(b)}
         \hfill}
     \vspace{0.42\textwidth}
     \caption{The corresponding relative percent error $E$ between the exact and numerical solutions, for Case (a) and (b), respectively.
 Contours are drawn and labeled  at levels, 1, 5, 10, 20, and 30\%.}\label{fig:err003}
 \end{figure}
We also quantify the error by calculating the relative percent
error between the exact and numerical solutions, defined as:
\begin{equation}
E=\frac{|\psi -\Psi|}{\langle|\Psi|\rangle}\times 100\%,
\label{eq:E}
\end{equation} after interpolating the numerical solution $\psi$
(obtained on a set of $(r,\theta)$ grid) onto the set of $RZ$ grid
on which the exact solution is defined. The corresponding results
are shown in Figure~\ref{fig:err003} (a) and (b), respectively, in
terms of contour plots of $E$ at certain levels between 1\% and
30\%. The overall pattern is that surrounding the initial line,
i.e., the spacecraft path at $Z=0$ in these cases, the errors are
generally small, especially for case (a), mostly $<5\%$, to
greater vertical extent. The errors increase with increasing
distance away from the initial line and toward corners of the
computational domain. In case (b), the performance of the solver
in the lower half domain ($Z<0$) is comparable to that in case
(a), although that in the upper half domain is much worse.

\begin{figure}
 \centerline{\includegraphics[width=0.47\textwidth,clip=]{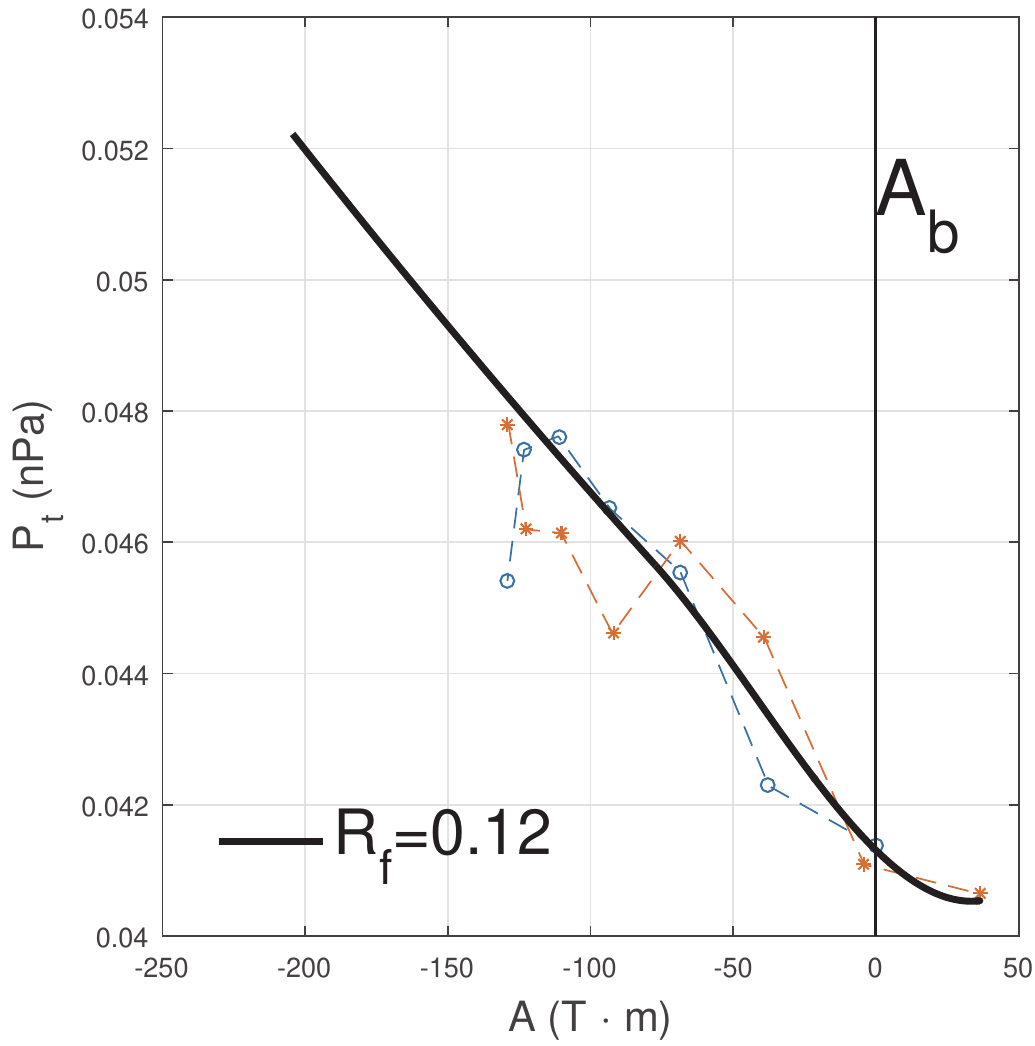}
 \includegraphics[width=0.53\textwidth,clip=]{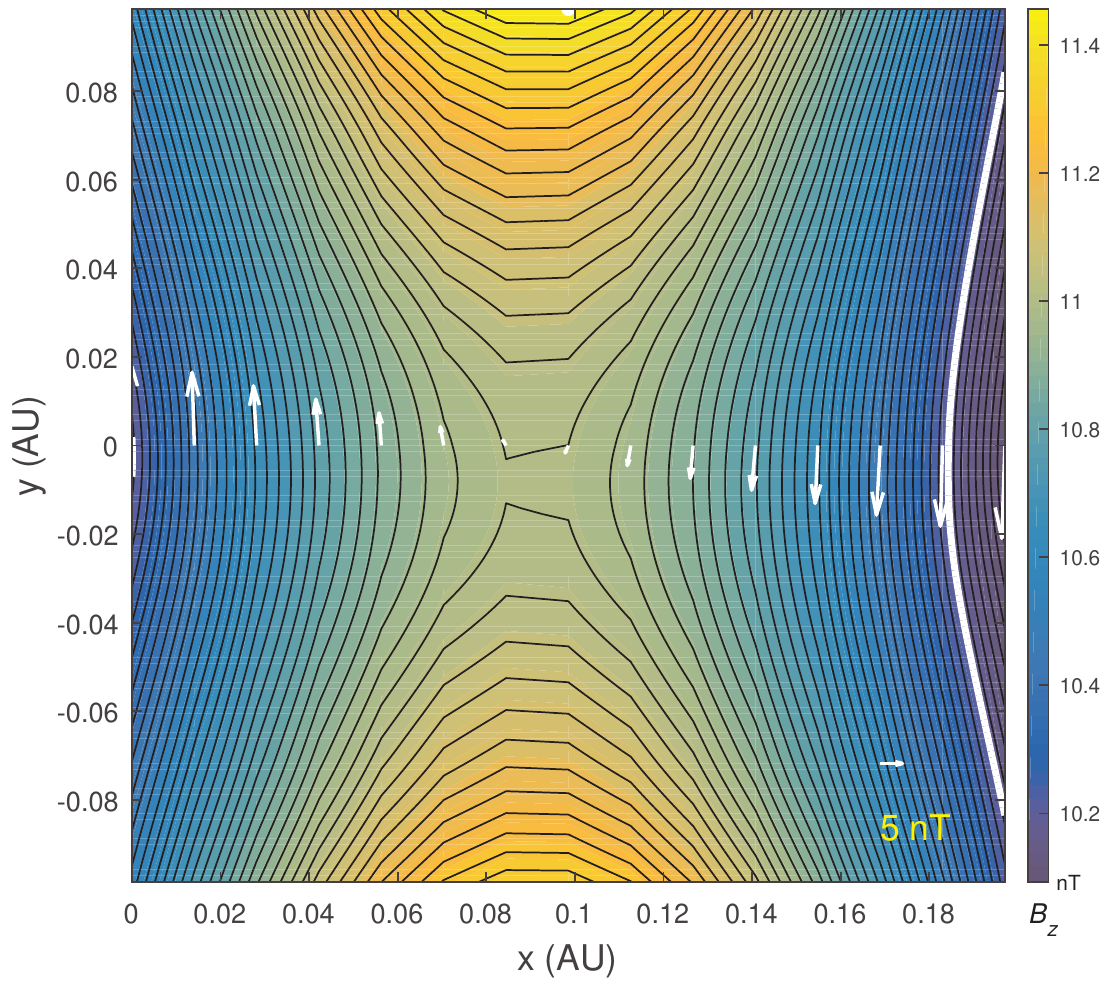}}
         \vspace{-0.47\textwidth}   
     \centerline{ \bf     
      \hspace{0.06 \textwidth} {(a)}
      \hspace{0.40\textwidth}  \color{white}{(b)}
         \hfill}
     \vspace{0.47\textwidth}
      \caption{(a) The  field-line invariant $P_t$ versus the flux function $A$ from the straight-cylinder GS reconstruction of case (a).
       (b) The corresponding cross section
       map.
 Formats are the same as Figures~\ref{fig:pta003} and \ref{fig:map003}, respectively. }\label{fig:GS0}
 \end{figure}
{\textbf{We also supply the time-series data from case (a) to the
standard straight-cylinder GS solver to check the effect of the
toroidal geometry and the specific magnetic field profile in this
case. The axial orientation is determined as $z=[-0.1710, 0.9838,
0.05440]$, in the $r_{sc}tn$ coordinate, primarily along $t$ (or
$\phi$) direction, in this case. The corresponding field-line
invariant $P_t=p+B_z^2/2\mu_0$ versus the flux function $A$ and
the functional fitting is shown in Figure~\ref{fig:GS0}a, yielding
a fitting residue $R_f=0.12$ of acceptable quality. The
reconstruction result, however, fails to yield a flux rope
solution, as shown in Figure~\ref{fig:GS0}b. It shows an X-line
type geometry, rather than an O-line type, i.e., that of a two and
a half dimensional magnetic flux rope (or island). This is due to
the peculiar magnetic field profile in this case (see
Figure~\ref{fig:B003}a), where the magnetic field magnitude
decreases significantly toward the center, down by about a half,
resulting in such a configuration of an X-line with much weaker
field strength in the middle. }}

 \begin{table}
 \caption{Comparison of the Outputs of the Numerical GS Solver with the Exact Solution (${\tt{NL}}=0.0)$ for
 $\theta_0=0$ (first 3 rows),
 \textbf{and $\theta_0=10^\circ$ (last row).}}\label{tbl:solver}
 \begin{tabular}{ccccc}
 \hline
$\tt{NL}$ & $R_f$ & $B_{\phi,max}$ (nT) & $\langle E\rangle$ &
$\Phi_p$ ($10^{12}$Wb/radian) \\
\hline
0.0 & - & 11.3 & - & 22.5 \\
0.01 & 0.10 & 11.3 & 5.5\% & 22.4 \\
0.1 & 0.28 & 10.5 & 9.5\% & 23.4 \\
0.01 & 0.10 & 11.3 & 5.2\% & 23.1 \\\hline
 \end{tabular}
 \end{table}
In summary, the various quantities derived from the toroidal GS
solutions are given in Table~\ref{tbl:solver}, whereas the
straight-cylinder GS solver fails to yield the flux rope solution.
As discussed above, case (b) generally exhibits more significant
errors than case (a), not surprisingly, due to its higher level of
noise, while case (a) yields fairly accurate results in this
limited set of outputs. Overall the errors in these quantities are
limited within 10\%, with the case (b) outputs approaching the
limit, which likely represents an extreme-case scenario.

\begin{figure}
 \centerline{\includegraphics[width=0.5\textwidth,clip=]{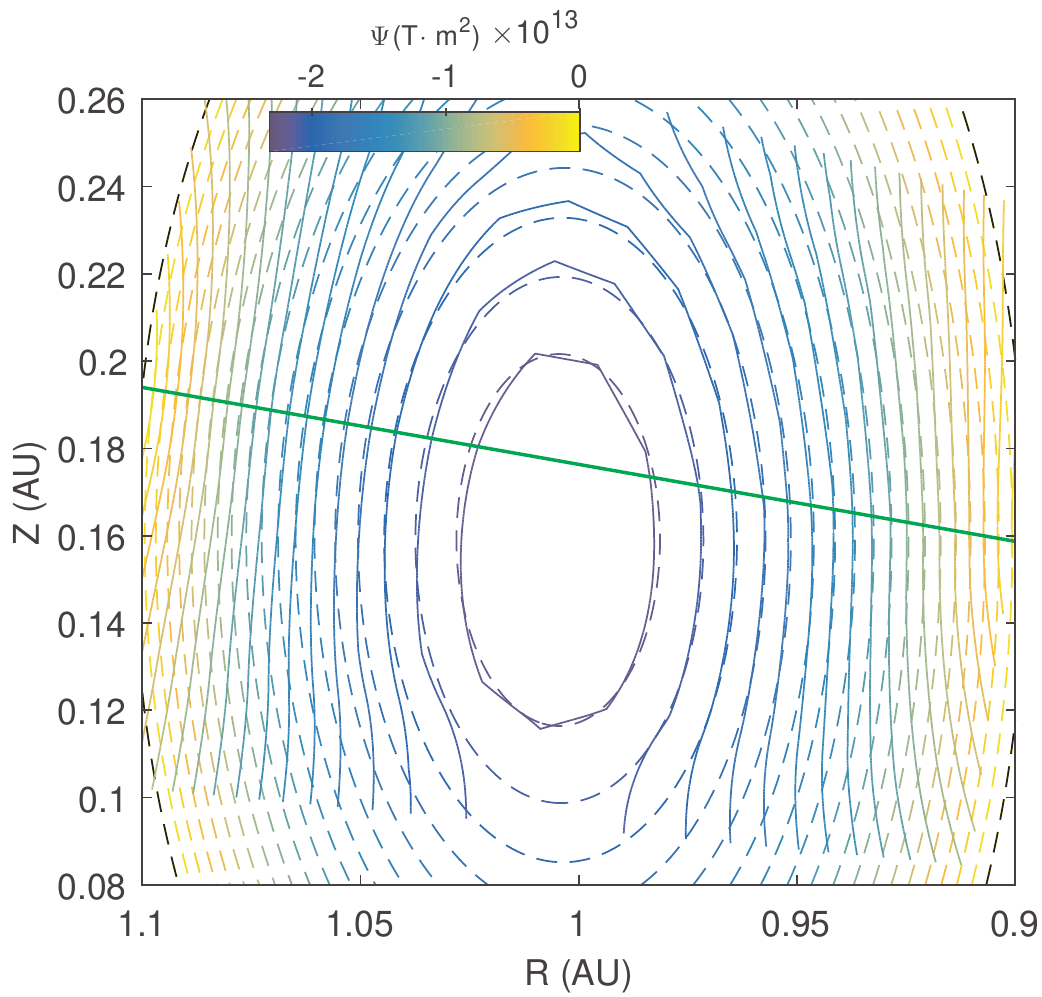}
 \includegraphics[width=0.5\textwidth,clip=]{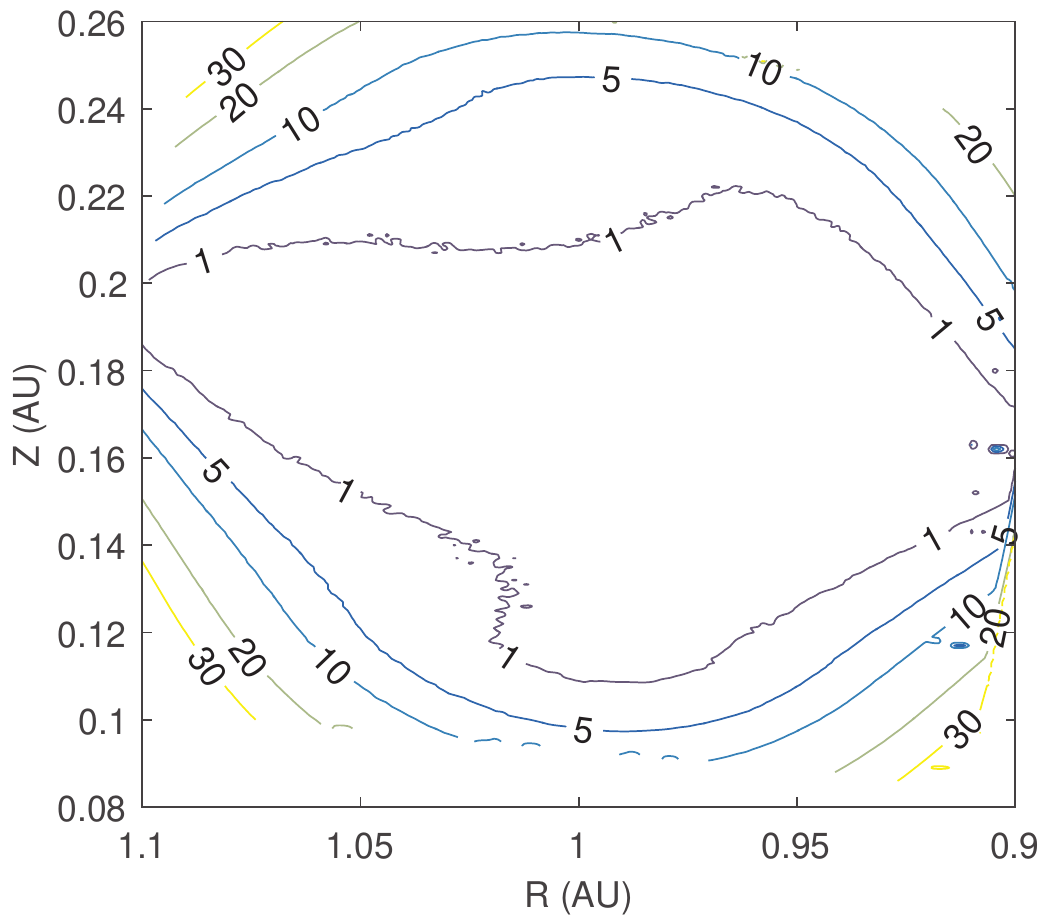}}
         \vspace{-0.48\textwidth}   
     \centerline{ \bf     
      \hspace{0.05 \textwidth} {(a)}
      \hspace{0.45\textwidth}  {(b)}
         \hfill}
     \vspace{0.48\textwidth}
      \caption{(a) The  exact  and  numerical solutions for $\theta_0=10^\circ$
       and ${\tt{NL}}=0.01$. (b) The corresponding contour plot of
       $E$.
 Formats are the same as Figures~\ref{fig:psi003} and \ref{fig:err003}, respectively. }\label{fig:psi010}
 \end{figure}
{\textbf{In addition, we also examine a case of
$\theta_0=10^\circ$ for ${\tt{NL}}=0.01$,  as one example of
nonzero $\theta_0$, such that the spacecraft is crossing along a
slanted path. Figure~\ref{fig:psi010} shows the comparison of
exact and numerical solutions, and the corresponding error
evaluation by the quantity $E$. The results are similar to the
case of $\theta_0=0$ of the same noise level. Because the
underlying numerical scheme is exactly the same as laid out in the
Appendix~\ref{app:solver}, the computation is still limited within
an annular region. The corresponding set of outputs is also listed
in Table~\ref{tbl:solver} (last row), for which the exact value of
$\Phi_p$ is 23.2 TWb/radian due to a slightly different boundary.
}}

\section{Conclusions and Discussion}\label{sec:summ}
In conclusions, we have developed a practical approach for
Grad-Shafranov (GS) reconstruction of magnetic flux ropes in
toroidal geometry, i.e., that of ring-shaped structures of
rotational symmetry. We devised a recipe to derive the unknown
geometrical parameters, i.e., the orientation of the rotation axis
$Z$ and the major radius of the torus $r_0$, from in-situ
spacecraft data and the toroidal GS equation. The algorithm
utilizes uncertainty estimates associated with the spacecraft
measurements to carry out proper $\chi^2$ minimization of the
deviation between the measured magnetic field components and GS
model outputs. Benchmark studies with analytic solutions to the GS
equation and added noise of known variances were carried out and
are presented to illustrate the procedures and to show the
performance of the numerical GS solver in the toroidal geometry.
Although shown separately and still  limited,  the results
indicate an absolute error of 9$^\circ$ in $Z$ axis orientation,
and a relative error of about 22\% for the major radius in one
case, while the relative percent errors in numerical GS solutions
are generally less than 10\%. {\textbf{The straight-cylinder GS
solver failed to yield the flux-rope solution for this particular
case.}}

We also make the computer codes written in Matlab publicly
available, accompanying this publication, which can also be
downloaded from the shared Dropbox
folder
\footnote{\url{https://www.dropbox.com/sh/wd5btkbldu5xvga/AABHQjCRRUH1NpEprmnKsccOa?dl=0}}.
The codes can generate most of the results presented in the main
text, and are also ready for applications to real events.
{\textbf{The included Readme file outlines the command-line
execution of the codes in Matlab to generate the results presented
here with little need to modify the codes. }} We encourage the
potential users to run the codes and to communicate with the
author on any issues that may arise.

We will present additional and more comprehensive benchmark
studies in a follow-up presentation, together with examples of
applications to real events \citep{2015ASPCH}. The limitation of
the current study is somewhat idealized conditions including
adding the artificial noise of normal distributions. The best
approach to overcome this might be to perform a more complete
benchmark study by utilizing the numerical simulation data, for
example, that of \citet{2004JASTPR}, where a toroidal flux rope
was propagated to 1 AU with synthetic data taken along two
separate spacecraft paths across the structure. Those data were
utilized in assessing the cylindrical flux rope models, and will
be re-examined by the current toroidal GS model. A more
comprehensive benchmark study combining Sections~\ref{subsec:Z}
and \ref{subsec:solver} will be presented.

The current implementation relies on the availability of reliable
estimate of measurement uncertainties, for example, associated
with magnetic field, which were usually derived from the
corresponding  higher resolution data. The utilization of such
uncertainty estimates in real events will be further investigated,
especially by using multiple time-series data from multiple
spacecraft across the same structure. As demonstrated in the
benchmark studies here, the contour of reduced $\chi^2\approx 1$
outlines the extent of uncertainties in GS model output. A more
complete assessment of such uncertainties associated with various output
parameters of the GS reconstruction results will be carried out in
the forthcoming study.

%

%

%
 \appendix

\section{Calculation of $R$ for a Given $Z$ at $O'$}\label{app:R}
We present one approach here the calculation of the array $R$ for
each point denoted by a vector $\mathbf{r}_{sc}$ along the
spacecraft path across the torus, for a given $Z$ axis of
components $(Z_r,Z_t,Z_n)$ at location $O'$, as illustrated in
Figure~\ref{fig:RSZ}. This is the distance between the origin $O$,
given by the vector $\mathbf{O}$ and $\mathbf{r}_{sc}$ (note all
vectors are given in the ${r}_{sc}tn$ coordinate):
\begin{equation}
R=|\mathbf{r}_{sc} - \mathbf{O}|.\label{eq:R}\end{equation} Then
the key step is to derive $\mathbf{O}$ for each $\mathbf{r}_{sc}$,
realizing that it is changing along $Z$ except for $Z$ being
perpendicular to $\mathbf{r}_{sc}$. It is trivial for the special
case when all
 $O$s coincide with one point along $Z$ (becoming $O'$ when $Z$ is
 perpendicular to the $r_{sc}t$ plane). So the following is for
 a
 general case and for $Z_t\ne 0$.

 From the known fact that both $O$ and $O'$, denoted by vector components $(r_o,t_o,n_o)$ and
 $(r',t',n')$, respectively, are along $Z$, it follows
 $$ \frac{r'-r_o}{Z_r}=\frac{t'-t_o}{Z_t}=\frac{n'-n_o}{Z_n}.$$
For $Z_t\ne 0$, we  obtain
\begin{equation}r_o=r' - \frac{Z_r}{Z_t}(t'-t_o) \label{eq:ro}\end{equation} and
\begin{equation}n_o=n' - \frac{Z_n}{Z_t}(t'-t_o).\label{eq:no}\end{equation}
By substituting them into $(\mathbf{r}_{sc} -\mathbf{O})\cdot
\hat{Z}=0$ and rearranging the terms, we obtain
\begin{equation}
t_o=\frac{(\mathbf{r}_{sc}
-\mathbf{r}_{op})\cdot{\hat{Z}}}{|Z|^2/Z_t} + t', \label{eq:to}
\end{equation}
where quantities on the right-hand side are all known with
$\mathbf{r}_{op}=(r',t',n')$. Then the vector $\mathbf{O}$ is
fully determined from equations~(\ref{eq:ro}) and (\ref{eq:no})
above. So is the array of $R$ from equation~(\ref{eq:R}) along the
spacecraft path.

Similar set of formulas can be obtained for the cases of $Z_r\ne
0$ or $Z_n\ne 0$.

\section{The Numerical GS Solver}\label{app:solver}
The numerical GS solver for the toroidal GS reconstruction is in
direct analogy to the straight-cylinder case \citep[see,
e.g.,][]{1999JGRH}, i.e., the approach by the Taylor expansion,
utilizing the GS equation~(\ref{eq:GSrth}) for evaluating the
2nd-order derivative in $\theta$.

To lay out the implementation of the numerical scheme in the code,
we denote $u_i^j=\Psi$ and $v_i^j=B_r$, where the indices $i$ and
$j$ represent uniform grids along dimensions $r$ and $\theta$,
with grid sizes $h$ and $\Delta\theta$, respectively. It is set
$\Delta\theta=0.01h$, and $\theta^j=(j-j_0)\Delta\theta +\theta_0$
($j=1:n_y$), where the index of the grid at $\theta=\theta_0$,
i.e., along the projected spacecraft path, is denoted $j_0$.
Changing $j_0$ will allow the spacecraft path where the initial
data are derived to shift away from the center line of the
computational domain. Then the solutions to the GS equation can
be obtained through usual Taylor expansions in $\theta$ (truncated
at the 2nd-order term with respect to $\Psi$), both upward and
downward from the initial line ($\theta=\theta_0$). For example,
for the upper half annular region $j\ge j_0$, noting the relations
$\frac{\partial \Psi}{\partial\theta}=rRB_r$, $\frac{\partial
\Psi}{\partial r}=RB_\theta$, and $R=R_0+r\cos\theta$, we obtain
(further denoting $rhs=-FF'$, as a known function of $u$ via the
functional fitting $F(\Psi)$, e.g., see Figure~\ref{fig:pta003}):
\begin{eqnarray}
u_i^{j+1}&=&u_i^j+(-v_i^j r_i R_i)\Delta\theta +
\frac{1}{2}a_i^j\Delta\theta^2 r_i^2,\\
v_i^{j+1}&=&v_i^j+\Delta\theta\left(-a_i^j\frac{r_i}{R_i}+\frac{r_i\sin\theta^j
v_i^j}{R_i}\right),
\end{eqnarray}
where the term $a_i^j$ involves the 2nd-order derivative in
$\theta$ and is evaluated via the GS equation,
$$a_i^j=rhs_i^j-\left(\frac{\partial^2 u}{\partial
r^2}\right)_i^j+\sin\theta^j v_i^j
-\left(\frac{1}{r_i}-\frac{\cos\theta^j}{R_i}\right)\left(\frac{\partial
u}{\partial r}\right)_i^j.$$ As usual, the partial derivatives in
$r$ are evaluated by 2nd-order centered finite difference for
inner grid points and one-sided finite difference for boundary
points.

Also similar to the usual straight-cylinder case, smoothing of the
solution at each step is necessary to suppress the growth of
numerical error. The same scheme is applied as follows to inner
grid points only \citep{2001PhDT........73H,2002JGRAHu} and for
the upper half domain ($j\ge j_0$):
$$\tilde{u}_i^j=\frac{1}{3}[k_1 u_{i+1}^j+k_2 u_i^j+k_3
u_{i-1}^j],$$ where the coefficients are $k_1=k_3=f_y$, and
$k_2=3-2f_y$, with $$f_y=\min\left\{0.7,
\frac{\theta^j-\theta_0}{\theta^{n_y}-\theta_0}\right\}.$$ The
same applies to $v$, and similarly to the lower half domain.

\section{The Hodograms for the Cases of  Submerged Spacecraft
Paths}\label{app:hodo}

These are the cases that cannot be dealt with by the toroidal GS
reconstruction technique developed here. These had been
traditionally analyzed by a fitting method to fit the spacecraft
measurements along its embedded path to a theoretical toroidal
flux rope model \citep[see., e.g.,][]{2015SoPh..290.1371M}. As we
discussed earlier and demonstrate further below, the ``projected"
spacecraft path takes a peculiar shape and the measured magnetic
field components possess certain features as indicated by the
associated hodogram pairs obtained from the usual minimum variance
analysis \citep{1998ISSIRS}.

\begin{figure}
 \centerline{\includegraphics[width=0.5\textwidth,clip=]{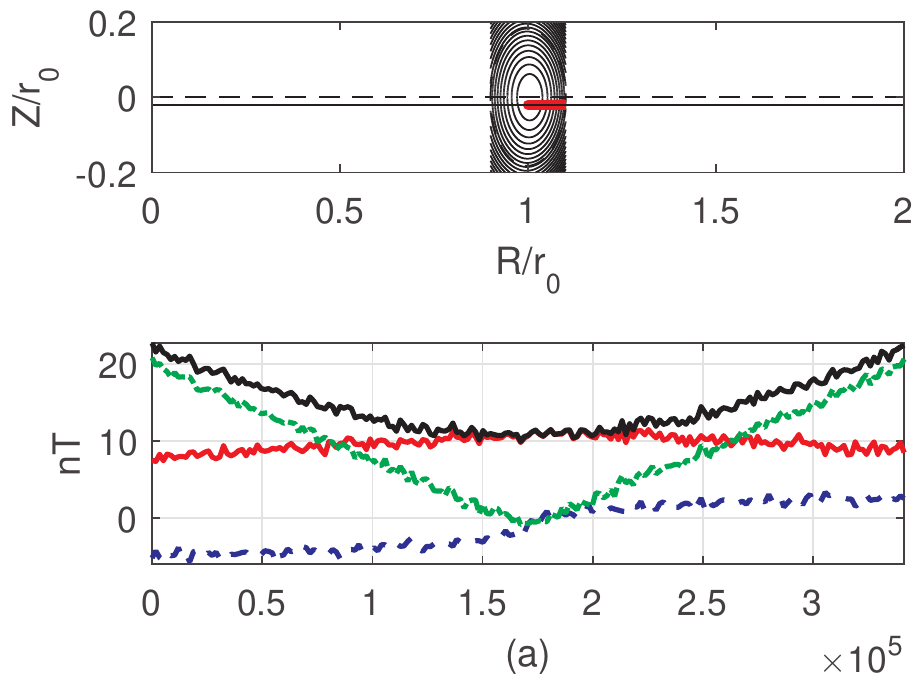}
 \includegraphics[width=0.5\textwidth,clip=]{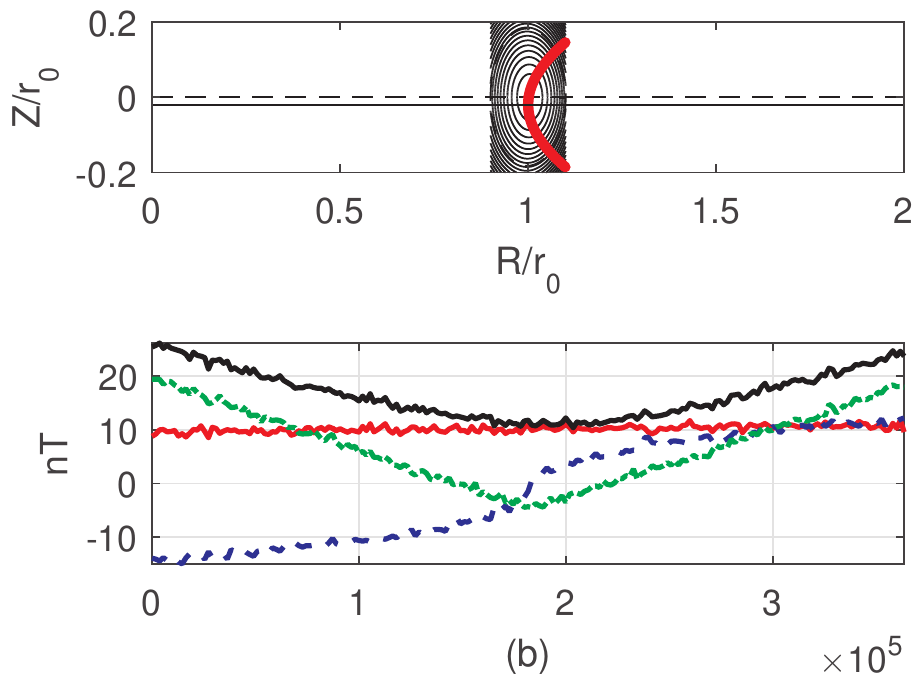}}
 \caption{The cases of submerged spacecraft paths: (a) a path perpendicular to $Z$, and (b) a slanted path. In each subplot,
 the upper panel shows the analytic solution and the projected spacecraft path in red in the same format as Figure~\ref{fig:psi332}, while
 the lower panel shows the magnetic field components along such a path (see legend of Figure~\ref{fig:B003}).}\label{fig:embedded}
 \end{figure}
\begin{figure}
 \centerline{\includegraphics[width=0.5\textwidth,clip=]{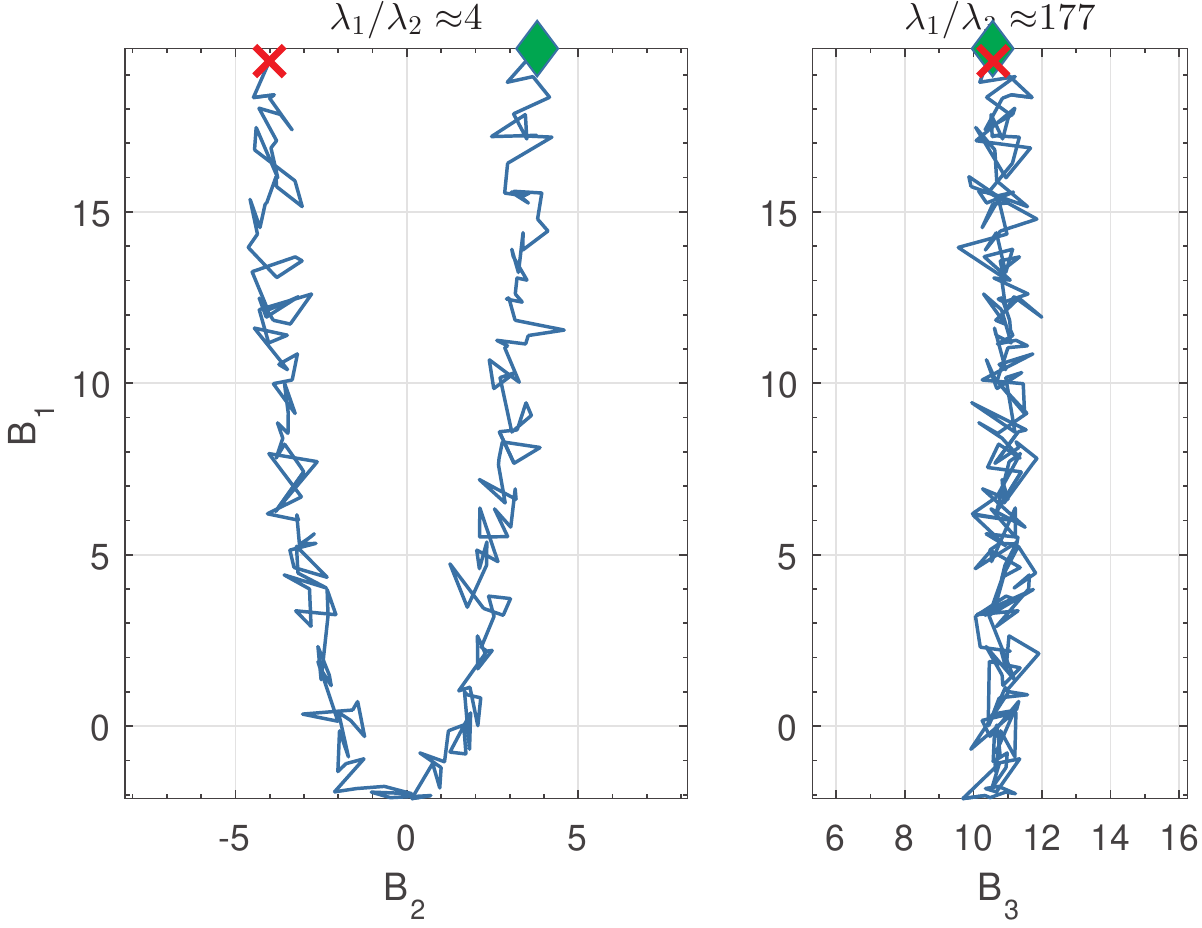}
 \includegraphics[width=0.5\textwidth,clip=]{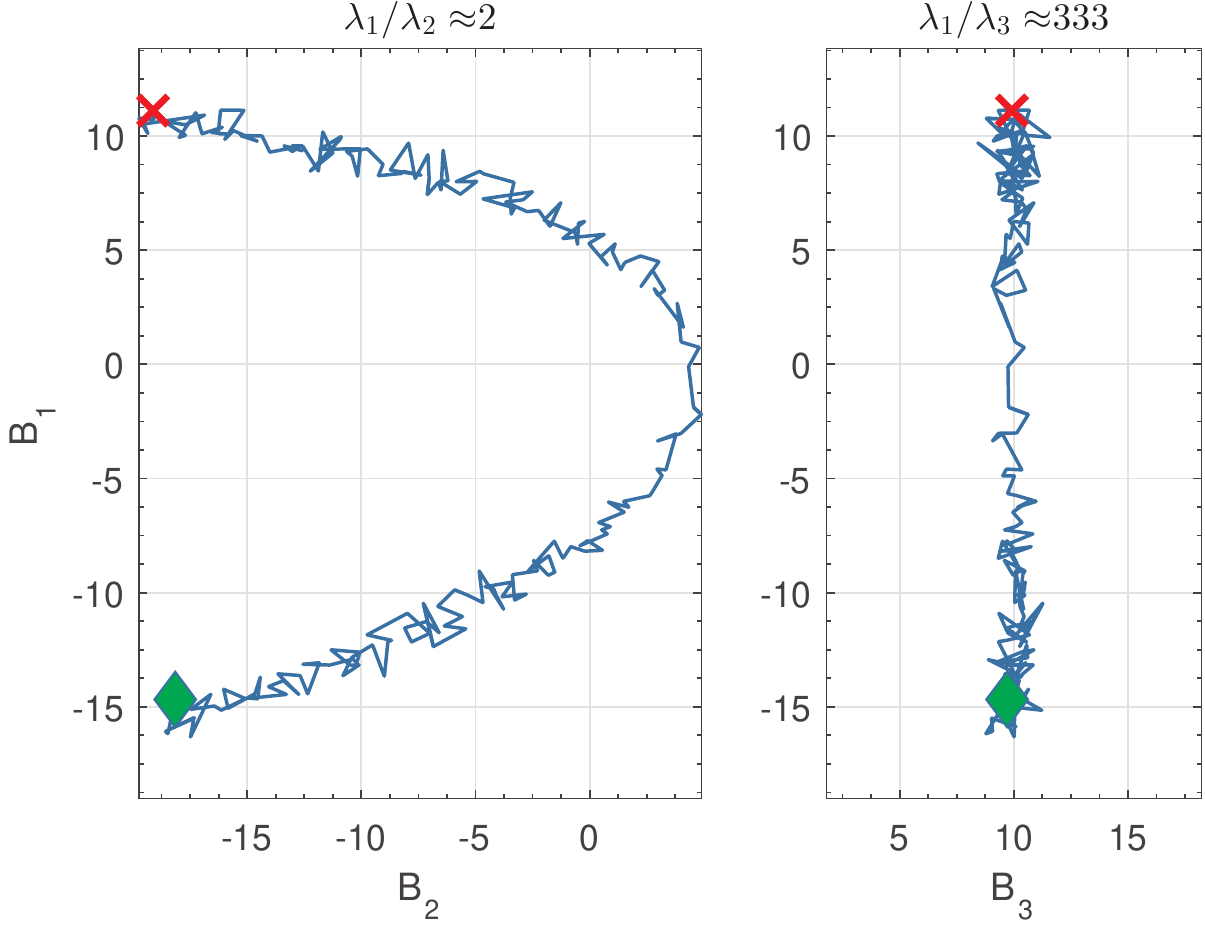}}
          \vspace{-0.41\textwidth}   
     \centerline{ \bf     
      \hspace{0.05 \textwidth} {(a)}
      \hspace{0.45\textwidth}  {(b)}
         \hfill}
     \vspace{0.41\textwidth}
 \caption{The hodogram pairs for Case (a) and (b), respectively. The magnetic field components are projected onto
 the maximum, intermediate, and minimum variance directions, $B_1$, $B_2$, and $B_3$, respectively, with corresponding
 eigenvalues, $\lambda_1$, $\lambda_2$, and $\lambda_3$. The diamond and cross symbols mark the beginning and end of the data interval.}\label{fig:hodos}
 \end{figure}

We again demonstrate these cases by utilizing the analytic
solutions presented in Section~\ref{sec:bench}. However here the
spacecraft path is specially taken, not to exit into the ``hole"
of the torus, but to be along the green line in
Figure~\ref{fig:torcoord}. Two such cases are presented in
Figure~\ref{fig:embedded}: (a) the spacecraft path is
perpendicular to $Z$ so that the ``projected" path is
double-folded onto itself, resulting in a situation where the
spacecraft is entering and exiting the cross section along the
same path but is only half-way through, and (b) the spacecraft
path is traversing along a slanted path, resulting in a warped
non-overlapping path across about half of the cross section. For
both cases, the magnetic field components change in time and show
clear features of symmetry or anti-symmetry, and possess
significant radial components, persistently $\sim 10$ nT
throughout the intervals. This is because that the spacecraft is
nearly encountering the same set of field lines during its inbound
and outbound passages, and of the up-down symmetry in these cases.
These features are clearly demonstrated by the corresponding
hodogram pairs shown in Figure~\ref{fig:hodos}. Especially in Case
(a), the $B_1$ versus $B_2$ hodogram exhibits a nearly closed loop
while the other one is double-folded, due to completely folded
path. Case (b) also displays significant rotation in $B_1$, about
180 degrees. It is worth noting that this type of pattern in Case
(a) is rarely reported in in-situ magnetic field measurements,
except for the case of \citet{2003GeoRL..30.2065R} where a nearly
360 degree rotation in the magnetic field was seen in the MC
interval. In other words, we caution that for this type of
configuration of a glancing pass by a spacecraft through a torus,
the magnetic field signatures as demonstrated here need to be
considered for proper modeling of these configurations.

The current implementation of the numerical GS solver cannot solve
for a solution over a significant portion of the cross section
because the ``projected" spacecraft path is no longer along a
single constant coordinate dimension, i.e., that of
$\theta\approx\theta_0=const$, across the whole cross-sectional
domain. A word of caution is that when interpreting the measured
time series in the $r_{sc}tn$ coordinate, they have to be taken
along the actual spacecraft path $\mathbf{r}_{sc}$ shown in
Figure~\ref{fig:torcoord}, not the ``projected" ones on the $RZ$
plane shown in Figure~\ref{fig:embedded}. Another important
observation from these preliminary analysis is that the field
rotation is actually more significant as indicated by the hodogram
pairs in these cases of ``glancing" passage of the spacecraft,
contrary to general perceptions one may have. Although this
provides proof of merits of flux rope model fitting to in-situ
spacecraft data under the toroidal geometry,  we urge that such
fitting better be done in the way of equation~(\ref{eq:chi2}) with
the mathematical rigor of proper uncertainty estimates for
quantitative and more objective assessment of the goodness-of-fit.

%
 \begin{acks}
QH acknowledges partial support from  NASA grants NNX14AF41G,
NNX12AH50G, and NRL contract N00173-14-1-G006 (funded by NASA LWS
under ROSES NNH13ZDA001N). The author benefits greatly from
decade-long collaboration with Prof.~Jiong Qiu. The author also
acknowledges illuminating discussions with the LWS FST team
members on flux ropes, in particular, Drs.~M.~Linton,
T.~Nieves-Chinchilla, B.~Wood, and the PSI group. The author is
also grateful for a few site visits to NRL hosted by
Dr.~M.~Linton.

 \end{acks}

%
%
 \bibliographystyle{spr-mp-sola}
 \bibliography{ref_master3}
%
%
%
%

\end{article}
\end{document}